\documentclass{JHEP3} 
\def\nn{\nonumber}

\def\eps{\varepsilon}
\def\l{\left}
\def\r{\right}

\def\beq{\begin{equation}}
\def\eeq{\end{equation}}
\def\bea{\begin{eqnarray}}
\def\eea{\end{eqnarray}}
\def\barr{\begin{array}}
\def\earr{\end{array}}
\def\be{\begin{equation}}
\def\ee{\end{equation}}
\def\bc{\begin{center}}
\def\ec{\end{center}}
\def\dd{\displaystyle}
\def\ov{\overline}
\def\cL{{\cal L}}
\def\Ne#1{{\cal N}=#1}
\def\De#1{{\it D}=#1}
\def\Zd{${\cal Z}_2$}
\def\eq#1{{eq.~(\ref{#1})}}

\def\op{{\mathcal{O}}}

\def\id{{\bf 1}}
\title{The minimal \boldmath{$\Ne4$} no-scale model \\ 
           from generalized dimensional reduction}
\author{Giovanni Villadoro and Fabio Zwirner \\
Theory Division, Physics Department, CERN, 
CH-1211 Geneva 23, Switzerland, \\ 
Dipartimento di Fisica, Universit\`a 
di Roma ``La Sapienza'', and \\ 
INFN, Sezione di Roma, P.le A. Moro 2, I-00185 Rome, Italy \\
E-mail: \email{giovanni.villadoro@roma1.infn.it}, 
\email{fabio.zwirner@roma1.infn.it}}
\preprint{CERN-PH-TH/2004-109 \\ ROMA-1376/04} 
\abstract{We consider the generalized dimensional reduction of pure
ungauged $\Ne4$, $\De5$ supergravity, where supersymmetry is
spontaneously broken to $\Ne2$ or $\Ne0$ with identically vanishing
scalar potential. We explicitly construct the resulting gauged $\De4$
theory coupled to a single vector multiplet, which provides the
minimal $\Ne4$ realization of a no-scale model. We discuss its
relation with the standard classification of $\Ne4$ gaugings,
extensions to non-compact twists and to higher dimensions, the $\Ne2$
theories obtained via consistent \Zd \ orbifold projections and
prospects for further generalizations.}
\begin{document} 
\section{Introduction}
Generalized dimensional reductions of higher-dimensional supergravity
\cite{ss} and superstring theories \cite{ssstr} provide elegant and
efficient mechanisms for supersymmetry breaking. In short, the
periodicity conditions in the compact extra dimensions can be
`twisted' by an $R$-symmetry of the higher-dimensional action. This
induces four-dimensional gravitino mass terms and breaks the
supersymmetries that do not commute with the twist, without generating
a four-dimensional vacuum energy. Also M-theory and superstring
compactifications with branes, orientifolds and fluxes (for a review
and an extensive list of references, see e.g. \cite{frey}) can be
related with twisted tori by suitable dualities \cite{ineq,dh}.  The
above theoretical constructions can play a r\^ole in solving some
crucial problems of higher-dimensional supergravity and superstring
theories: vacuum selection, supersymmetry breaking, moduli
stabilization, generation of hierarchies. The corresponding $\De4$
effective theories are gauged extended (${\cal N} > 1$)
supergravities~\footnote{Throughout this paper, we shall count the
number of supersymmetries in any dimension in terms of the equivalent
number ${\cal N}$ of four-dimensional supersymmetries.}, where a
subgroup of the isometry group of the scalar manifold (which includes
the $R$-symmetry group) is promoted to a local invariance.

It has been known for quite some time that generalized dimensional
reductions of $D=11$ \cite{cj} or minimal $D=10$ \cite{ch}
supergravity give rise to gauged $\Ne8$ \cite{ss8} or $\Ne4$
\cite{ss4} $\De4$ supergravities, and that daughter theories with a
lower number of supersymmetries can be obtained by suitable orbifold
projections. Systematic investigations of the effective gauged $\De4$
theories have recently been performed starting from $\Ne8$
\cite{gau8a,gau8b} or $\Ne2$ \cite{gau2} supergravities; many
results have been obtained also for the $\Ne4$ theories
\cite{gau4a,gau4b} that are the focus of the present paper. Indeed,
even in the case of $\Ne1$ orbifolds of superstrings or M-theory, the
effective $\De4$ superpotential for the light (bulk) states coming
from the untwisted sector can be associated with the gauging of the
underlying $\Ne4$ theory \cite{dkz}.

The goal of the present paper is to consider generalized dimensional
reductions of pure $\Ne4$, $\De5$ ungauged supergravity \cite{cremmer,
awatow}, and to derive explicitly Lagrangian and transformation laws
for the resulting low-energy theory, a gauged $\Ne4$, $\De4$
supergravity coupled to a single vector multiplet. Such a theory can
be legitimately called the minimal $\Ne4$ no-scale model, in analogy
with the minimal no-scale models with $\Ne1$ \cite{noscale} and $\Ne2$
\cite{seven} constructed long ago. It is also the minimal $\Ne4$
no-scale model that allows for the partial breaking to $\Ne2$, whereas
partial breaking to $\Ne3$ or $\Ne1$ requires the presence of
additional vector multiplets \cite{tsozin}: we will see that the
minimal $\Ne2$ no-scale model with partial breaking \cite{fgp}, which
contains one vector multiplet and one hypermultiplet, corresponds to a
consistent \Zd \ orbifold projection of our minimal $\Ne4$ no-scale
model \cite{alten}. The choice of $\Ne4$ for our detailed study is
motivated not only by the need of filling a gap in the existing
literature, but also by the fact that $\Ne4$ is the minimal amount of
supersymmetry shared by all stable superstring theories, and their
corresponding supergravities, in $D=10$. Within $\Ne4$ theories, the
choice of $\De5$ as the starting point for the generalized dimensional
reduction allows for minimality (when no $\De5$ vector multiplets are
included) as well as for maximum generality (when an arbitrary number
of $\De5$ vector or tensor multiplets are included), and prepares the
ground for further discussions of the minimal Randall--Sundrum model
\cite{rs} with extended supersymmetry \cite{extrs}, which is based on
a pure gauged $\Ne4$, $\De5$ supergravity \cite{awatow,romans}.

The paper is organized as follows. In section~2 we briefly recall the
field content and the symmetries of pure, ungauged $\Ne4$, $\De5$
supergravity. In section~3 we derive the ungauged $\Ne4$, $\De4$
effective supergravity, coupled to a single vector multiplet, that
originates from standard dimensional reduction. In particular, we
identify the `electric' subgroup of the $\De4$ duality group that acts
linearly on the gauge potentials and on the scalar fields.  The gauged
$\Ne4$, $\De4$ supergravity, originating from the pure ungauged $\De5$
theory by generalized dimensional reduction, is derived in
section~4. Twisting the periodicity conditions by an $R$-symmetry
transformation, controlled by two independent parameters, does not
generate a potential, and leads to $\Ne4$ no-scale models with
spontaneous breaking of half or all of the supersymmetries. As already
stressed in similar contexts \cite{tsozin, ineq, gau8a, gau8b}, the
effective $\Ne4$, $\De4$ gauged supergravity does not fit the standard
classification \cite{wag}: the reason is that formulations of the
theory that are equivalent, in the ungauged case, via duality
transformations, are no longer equivalent in the gauged case. For
completeness, we also discuss the possibility of a non-compact twist,
which generates a positive-definite four-dimensional potential without
critical points \cite{ncgau}. In section~5, we explain how our results
can be applied to generalized dimensional reductions of pure $\Ne4$
supergravity from $D$ to $D-1$ dimensions, and we discuss the examples
of the non-chiral (2,2) and the chiral (4,0) theories reduced from six
to five dimensions. In the former case, we obtain new `flat' gaugings
of $\Ne4$ in $\De5$, not included in the classification of
\cite{dagata}, which was limited to gaugings of semisimple and abelian
groups. In the latter case, we generalize the flat gauging of
ref.~\cite{romans} to partial breaking. In section~6 we discuss
consistent \Zd \ orbifold truncations in the presence of the
Scherk-Schwarz twist, and the main features of their $\Ne2$, $\De4$
effective theories: we find total or partial supersymmetry breaking
with vanishing potential, in agreement with previous results
\cite{fgp, alten, gau2}. In the concluding section, we summarize again
our results and we comment on prospects for further work. Our
conventions are explicitly spelled out in appendix~A. The Lagrangian
and the transformation laws of pure ungauged $\Ne4$, $\De5$
supergravity are recalled in appendix~B. The details of the $\Ne4$,
$\De4$ Lagrangian and transformation laws, relevant to both standard
and generalized dimensional reduction, are collected in appendix~C.

%%%%%%%%%%%%%%%%%%%%%%%%%%%%%%%%%%%%%%%%%%%%%%%%%%%%%%%%%%%%%%%%%%%%
%%                          SECTION                               %%
%%%%%%%%%%%%%%%%%%%%%%%%%%%%%%%%%%%%%%%%%%%%%%%%%%%%%%%%%%%%%%%%%%%%

\section{\boldmath{$\Ne4$}, \boldmath{$\De5$} ungauged supergravity}

The field content of pure $\Ne4$, $\De5$ supergravity, whose ungauged
version was first constructed in \cite{cremmer,awatow}, is just the
gravitational multiplet: one graviton $g_{MN}$, four gravitinos
${\psi_M}_a$, five plus one vectors $V_M^i+v_M$, four spin-1/2
fermions $\chi_a$ and one real scalar $\phi$. The automorphism group
of the $\Ne4$, $\De5$ supersymmetry algebra ($R$-symmetry) is $USp(4)$.
Fields with indices $a=1,\dots,4$ and $i=1,\dots,5$ transform in the
$\bf 4$ and $\bf 5$ irreducible representations, respectively; all the
other fields are singlets. In a schematic notation that will be
convenient in the following, we summarize the content of the
gravitational multiplet as:
\beq
\label{shortnot}
_{\De5}[1,4,5+1,4,1]^{\Ne4}_{m=0} 
% \qquad \Leftrightarrow \qquad 
%_{\De5}[g_{MN},{\psi_M}_a,V_M^i+v_M,\chi_a,\phi]^{\Ne4}_{m=0} 
\, .
\eeq
The numbers in brackets count the representations of different `spin',
with the latter decreasing from left to right in steps of $1/2$. The
subscript `$_{m=0}$' recalls that we are dealing with a massless
multiplet. Further details on our conventions can be found in
appendix~\ref{app:conventions}. The Lagrangian and the supersymmetry
transformations are collected in appendix~\ref{app:5dlag}.

Before moving to the study of standard and generalized dimensional
reductions to $\De4$, it is useful to recall the local and global
symmetries of the $\De5$ theory. The local symmetries are general
coordinate transformations, $\Ne4$ supersymmetry and a $U(1)^6$ gauge
invariance, associated with the six vector fields and with respect to
which no fields are charged. The global symmetry of the theory, or
`$U$-duality' group, is in this case $USp(4) \times SO(1,1)$. It is an
electric subgroup of the $D=4$ $Sp(14,\bf R)$ duality group acting on
the space of the vector field strengths and their duals \cite{GZ}, in
the sense that it acts linearly on the vector potentials. The action
of $SO(1,1)$ on the fields is:
\beq \label{eq:so115d}
V_M^i \rightarrow e^{-\lambda} \ V_M^i \,,\qquad  v_M 
\rightarrow e^{2\lambda} \ v_M \,,\qquad  \phi \rightarrow 
\phi+\sqrt6\lambda \, , 
\eeq
while $USp(4) \sim SO(5)$ acts canonically on the $a$ and $i$ indices.

If we couple $\Ne4$, $\De5$ pure supergravity to $n$ vector multiplets,
the $U$-duality group is enlarged to $SO(5,n) \times SO(1,1)$, 
and the scalar manifold becomes:
\beq
\frac{SO(5,n)}{SO(5) \times SO(n)} \times SO(1,1) \, .
\eeq
In this case, the combination of generators used for the twist and
giving rise to a flat $\De4$ gauging can be embedded in the maximal
compact subgroup of $SO(5) \times SO(n)$.  However, switching on the
$SO(n)$ sector does not increase the possibilities for supersymmetry
breaking via generalized dimensional reduction, it just increases the
dimension of the $\De4$ gauge group~\footnote{More possibilities arise
only when considering $\Ne4$, $\De4$ gaugings \cite{tsozin, gau8a,
gau8b, gau4a, gau4b} that cannot be originated just by Scherk--Schwarz
reductions of $\Ne4$ theories, but require instead generalized
dimensional reductions of $\Ne8$ theories combined with a \Zd \
orbifold that explicitly breaks half of the
supersymmetries.}. Moreover, some of the resulting non-minimal $\Ne4$
no-scale models have already been constructed as truncations of gauged
$\Ne8$ theories \cite{gau4a, gau4b}. For these reasons, in the
following we will focus on the minimal model ($n=0$), corresponding to
pure $\Ne4$, $\De5$ supergravity.
 
%%%%%%%%%%%%%%%%%%%%%%%%%%%%%%%%%%%%%%%%%%%%%%%%%%%%%%%%%%%%%%%%%%%%
%%                          SECTION                               %%
%%%%%%%%%%%%%%%%%%%%%%%%%%%%%%%%%%%%%%%%%%%%%%%%%%%%%%%%%%%%%%%%%%%%

\section{Standard dimensional reduction}

We now describe the salient features of the standard dimensional
reduction of the $\Ne4$, $\De5$ theory. For this purpose, we consider
only the kinetic and Chern--Simons terms of the $\De5$
Lagrangian \eq{eq:5dN4lag}:
\bea
e_5^{-1}{\cal L}_5^{kin}&=& -R_5 -\frac12\partial_M \phi \, \partial^M
\phi -\frac14 X^4 v_{MN}v^{MN}-\frac14 X^{-2} V^i_{MN} V^{MN}_i \nn \\
&& -\frac18 e_5^{-1} \varepsilon^{MNRST} V_{MN}^i V_{RS}^i v_T
-\frac{i}{2} \ov \psi_M^a \gamma^{MNR} D_N \psi_{R \, a} +
\frac{i}{2} \ov\chi^a\gamma^M D_M \chi_a \, .
\label{l5kin}
\eea
The first step is to decompose the fields with five-dimensional
indices into fields with four-dimensional indices:
\begin{equation}
E^{\ A}_M = \l ( \begin{array}{c c} \rho^{-1/2}e^\alpha_\mu & \rho
A_\mu \\ 0 & \rho\end{array} \r ), \ {\psi_M}_a = \l (\barr{c}
{\psi_\mu}_a \\{\psi_y}_a\earr \r), \ V_M^i=\l (\barr{c} V_\mu^i \\
V_5^i\earr \r), \ v_M=\l (\barr{c} v_\mu \\ v_5 \earr \r) \, . \nn
\label{funf}
\end{equation}
We can then assume that the zero modes do not depend on the fifth
coordinate $y$ and perform the following field redefinitions:
\bea
&\psi_\mu^a = \rho^{-1/4} \eta_\mu^a + \l ( A_\mu +\frac{i}{2}
\rho^{-3/2} {\widehat \gamma} \gamma_\mu \r) \psi_y^a \,,& \nn \\
&{\psi_y}_a=\rho^{5/4} {{\psi}'_y}_a \,, \qquad
\chi_a=\rho^{1/4}\chi'_a \, , & \nn  \\ &V_\mu^i =
B_\mu^i + V_5^i A_\mu \, , \qquad v_\mu = b_\mu +v_5 A_\mu \,, &\nn \\
&t = \rho \, X^{-2} \, ,\qquad \tau = v_5 \, , \qquad \varphi_0 =
\sqrt{2} \, \rho \, X \, , \qquad \varphi^i = V_5^i \, . &
\label{eq:fieldredef}
\eea
The first two lines allow us to `ortho-normalize' the $\De4$ fermionic
kinetic terms, the third one avoids mixing terms of the form
`$V^{\mu\nu}A_\mu \partial_\nu \varphi$', and the last one makes the
duality invariance of the scalar sector manifest.

After moving from the field basis $(e_\mu^\alpha, \psi_\mu,
\psi_y, \chi, V_\mu^i, v_\mu, A_\mu, V_5^i, v_5, \rho, \phi)$ 
to the field basis ($e_\mu^\alpha, \eta_\mu, \psi_y^\prime, 
\chi', B_\mu^i, b_\mu, A_\mu, \varphi^i, \tau, \varphi_0, t$),
the part of the reduced Lagrangian coming from \eq{l5kin} reads:
\bea
e_4^{-1}\cL&=& -R_4- \frac12 \frac{ \partial_\mu t \partial^\mu t 
+ \partial_\mu \tau \partial^\mu \tau }{t^2}
- \frac{ \partial_\mu \varphi_0 \partial^\mu \varphi_0
+\partial_\mu \varphi_i\partial^\mu \varphi^i}{\varphi_0^2} 
\nn \\
&&- \frac14 \, t \, \l ( B_{\mu\nu}^i +
\varphi^i \, A_{\mu\nu} \r)^2 - \frac18 {\varphi_0^2 \over t} \l
( b_{\mu\nu} + \tau \, A_{\mu\nu} \r)^2 - {t \, \varphi_0^2
\over 8} A_{\mu\nu} A^{\mu\nu} \nn \\ && - \frac{\tau}{8} e_4^{-1}
\varepsilon^{\mu\nu\rho\sigma} \l ( B_{\mu\nu}^i + \varphi^i \,
A_{\mu\nu} \r) \l ( B_{\rho\sigma}^i + \varphi^i \, A_{\rho\sigma} \r)
\nn \\ && - \frac{\varphi^i}{4} e_4^{-1}
\varepsilon^{\mu\nu\rho\sigma} b_{\mu\nu} \l ( B_{\rho\sigma}^i +
\frac12 \varphi^i \, A_{\rho\sigma} \r)
\nn \\
&&-\frac{i}{2} \ov \eta_\mu^a \gamma^{\mu\nu\rho} D_\nu 
\eta_{\rho \, a} + \frac{i}{2} \ov \chi^a \gamma^\mu D_\mu \chi_a 
+\frac34 i \, \ov \psi^a_y \gamma^\mu D_\mu \psi_{y \, a} 
+ \dots \, , 
\label{eq:redlagkin0} 
\eea
where the dots stand for interaction terms and the primes in
\eq{eq:fieldredef} have been suppressed. The Lagrangian can be
rewritten more compactly as:
\bea
e_4^{-1}\cL&=& -R_4- \frac12 \frac{ \partial_\mu t \partial^\mu t 
+ \partial_\mu \tau \partial^\mu \tau }{t^2}
- \frac{ \partial_\mu \varphi_0 \partial^\mu \varphi_0
+\partial_\mu \varphi_i\partial^\mu \varphi^i}{\varphi_0^2} 
\nn \\
&&-\frac14  \, g_{IJ} \, F^I_{\mu\nu} F^{J \ \mu\nu}
-\frac18 \, e_4^{-1} \theta_{IJ} \, 
\varepsilon^{\mu\nu\rho\sigma} F^I_{\mu\nu} F^{J}_{\rho\sigma} 
\nn \\
&&-\frac{i}{2} \ov \eta_\mu^a \gamma^{\mu\nu\rho} D_\nu 
\eta_{\rho \, a} + \frac{i}{2} \ov \chi^a \gamma^\mu D_\mu \chi_a 
+\frac34 i \, \ov \psi^a_y \gamma^\mu D_\mu \psi_{y \, a} 
+ \dots \, , 
\label{eq:redlagkin}
\eea
where the field strengths $F^I_{\mu\nu}$ ($I=1,\dots,7$) are defined
through
\be
F^I_{\mu\nu} = 2 \, \partial_{[\mu} B^I_{\nu]} \, ,
\qquad
B_\mu^I = (B_\mu^i, \, b_\mu, \, A_\mu) \, ,
\ee
and the symmetric field-dependent matrices $g_{IJ}$ and $\theta_{IJ}$
are:
\be
g_{IJ} = \l ( \begin{array}{c c c} 
t \, \delta_{ij} & 0 & t \, \varphi_i \\
&& \\ 
0 & \frac{\dd \varphi_0^2}{\dd 2 \, t} &  
\frac{\dd \tau \, \varphi_0^2}{\dd 2 \, t}
\\ & & \\
t \, \varphi_j & 
\phantom{bla} 
\frac{\dd \tau \, \varphi_0^2 }{\dd 2 \, t} 
\phantom{bla}
& t \, \varphi_i \, \varphi^i +
\frac{\dd \tau^2 \, \varphi_0^2}{\dd 2 \, t}
+ \frac{\dd t \, \varphi_0^2 }{\dd 2}
\end{array} \r ) \, , \label{eq:defg}
\ee
\be
\theta_{IJ}=\l ( \begin{array}{c c c} 
\tau \, \delta_{ij} & \varphi_i  & \tau \, \varphi_i \\ 
& & \\
\varphi_j & 0 & \frac{\dd \varphi_i \, \varphi^i}{ \dd 2} \\ 
& & \\
\tau \, \varphi_j & 
\phantom{bla} \frac{\dd \varphi_i \, \varphi^i}{\dd 2} 
\phantom{bla}
& \tau \, \varphi_i \, \varphi^i \end{array} \r ) 
\, . 
\label{eq:deftheta}
\ee
Notice that the metric $g_{IJ}$, which controls the vector kinetic
terms, is a positive-definite and non-singular symmetric matrix for
all the allowed scalar field configurations ($\varphi_0>0$, $t>0$,
$\varphi_i \in {\mathbf R}$ and $\tau \in {\mathbf R}$).

The reduced theory is a $USp(4)$, $\Ne4$, $\De4$ supergravity coupled
to one vector multiplet.  In the schematic notation of \eq{shortnot},
we can write its spectrum as:
\beq
_{\De4}[1,4,6,4,2]^{\Ne4}_{m=0}
\quad
+
\quad
_{\De4}[0,0,1,4,6]^{\Ne4}_{m=0} \, .
\eeq
The eight scalar fields (one coming from the metric, six coming from
the vector fields and the $\De5$ dilaton) parametrize the coset
manifold:
\beq
\frac{SU(1,1)}{U(1)} \times \frac{SO(6,1)}{SO(6)} \, .
\eeq
The first factor is a K\"ahler manifold, parametrized by the complex
scalar
\be
T = t + i \, \tau \, ,
\qquad
(t, \, \tau \in {\mathbf R}) \, ,
\ee
with a K\"ahler potential
\be
K = - \log ( T + \ov{T} ) \, ,
\ee
which implies (remembering our convention for the Einstein term) the
kinetic term:
\be
- 2 \, K_{\ov{T}T} (\partial_\mu \ov{T}) 
(\partial^\mu T) = - \frac12 \frac{ (\partial_\mu t) 
(\partial^\mu t) + (\partial_\mu \tau) (\partial^\mu \tau)}{t^2} \, .
\ee
The second factor is the scalar manifold of the real fields
$(\varphi_0,\varphi_i)$.

The $U$-duality group of the reduced theory is $SU(1,1) \times
SO(6,1)$. It is associated with a solvable Lie algebra, which can be
decomposed as follows:
\bea
SU(1,1)&=& SO(1,1) + {\bf 1}^+ + {\bf 1}^- \, , \nn \\
SO(6,1)&=& SO(1,1)+SO(5)+{\bf 5}^+ +{\bf 5}^- \, . 
\eea
The $U$-duality group is not completely embedded in the electric
subgroup of the full duality group $Sp(14,{\mathbf R})$, acting on the
seven field strengths and their duals. As a result, only part of the
$U$-duality group can be gauged. In particular, the transformations
associated with the generators ${\bf 1}^-$ and ${\bf 5}^-$ do not have
a well-defined action on the elementary fields. As long as the theory
remains ungauged, however, $Sp(14,{\mathbf R})$ transformations can be
used to connect different equivalent formulations of the same theory:
besides the $USp(4)$ of this paper, known examples are the $SO(4)$ of
\cite{so4d4} and the $SU(4)$ of \cite{csf}.

The non-trivial global symmetries of the theory with a well-defined
action on the elementary fields are the $\bf 5^+$+$\bf 1^+$
translations, the two dilatations and the $SO(5) \sim USp(4)$
transformations.

The $\bf 5^+$+$\bf 1^+$ translations derive from the U(1)$^6$ gauge
symmetry of the $\De5$ theory. Their action on the `axions' is then
\beq
\tau \rightarrow \tau + \alpha_6 \,, \qquad \varphi_i \rightarrow
\varphi_i+\alpha_i \, . \nn
\eeq
The vectors transform as
\be
b_\mu \rightarrow b_\mu - \alpha_6 \, A_\mu \, , 
\qquad B_\mu^i \rightarrow B_\mu^i- \alpha_i \, A_\mu \, , 
\ee
where we used the definitions of $(b_\mu, \, B^i_\mu)$ and the fact
that $(v_\mu, \, V^i_\mu)$ do not transform. The scalar sector is
explicitly invariant. Since
\be 
\delta \, (b_{\mu \nu} + \tau \, A_{\mu \nu} ) = 0 \, , \qquad
\delta \, (B_{\mu \nu}^i + \varphi^i \, A_{\mu \nu} ) = 0 \, , 
\ee
it is easily shown that also the other terms are invariant, up to
total derivatives coming from the Chern--Simons term~\footnote{As will
be clear later, this is the reason why the gauging of shift symmetries
requires extra Chern--Simons terms to be added to the $\De4$
Lagrangian.}.

The two dilatations act on the scalar fields as follows:
\beq
t \rightarrow \beta t \, ,
\quad 
\tau\rightarrow \beta \tau \, , 
\quad
\varphi_0\rightarrow \gamma \varphi_0 \, ,
\quad \varphi_i\rightarrow \gamma \varphi_i \, , 
\eeq
and the invariance of the scalar sector is manifest. The invariance of
the whole Lagrangian is obtained by requiring that the vector fields
transform as
\be
B_\mu^i \rightarrow \beta^{-1/2} \, B_\mu^i \, , 
\quad 
b_\mu \rightarrow \beta^{1/2} \, \gamma^{-1} \, b_\mu \, , 
\quad 
A_\mu \rightarrow \beta^{-1/2} \, \gamma^{-1} \, A_\mu \, .
\ee
Notice that the double $SO(1,1)$ is a symmetry of only the reduced
theory, and it is not valid when the Kaluza--Klein modes are retained,
unless $\gamma = \beta^{-1/2}$: in this case, the five-dimensional
$SO(1,1)$ symmetry of \eq{eq:so115d} is recovered.

Finally, the invariance under the $SO(5) \sim USp(4)$ symmetry is
manifest.

Summarizing, the reduced action is invariant under the global symmetry
$\{ [ USp(4) \times SO(1,1)] \circledS {\cal T}^5 \}$ $\times$ $[
SO(1,1) \circledS {\cal T} ]$, with a semidirect product structure,
and under the local $U(1)^7$ group, with respect to which, however, no
field is charged.
%

%%%%%%%%%%%%%%%%%%%%%%%%%%%%%%%%%%%%%%%%%%%%%%%%%%%%%%%%%%%%%%%%%%%%
%%                          SECTION                               %%
%%%%%%%%%%%%%%%%%%%%%%%%%%%%%%%%%%%%%%%%%%%%%%%%%%%%%%%%%%%%%%%%%%%%

\section{Generalized dimensional reduction} 
\label{sec:SSred}

In the previous section we have identified the $\Ne4$ supergravity,
coupled to one vector multiplet, obtained from the pure ungauged
$\De5$ theory by standard dimensional reduction. In particular, we
have observed that the ungauged theory does not have the form of the
known $SO(4)$ and $SU(4)$ theories, but it is equivalent to them via a
duality transformation. However, as argued in \cite{tsozin, ineq,
gau8a, gau8b}, inequivalent formulations of gauged supergravities can
be obtained by considering different embeddings of the $U$-duality group
in the full duality group $Sp(14,\bf R)$. We can then use the
Scherk--Schwarz mechanism to obtain a new $\Ne4$, $\De4$ gauged
supergravity. We are going to find, in analogy with \cite{gau8a}, two
remarkable properties. First, if the twist belongs to $USp(4)$ the
supergravity gauging corresponds to a flat but non semisimple group,
which guarantees a vanishing scalar potential at the
minimum. Moreover, a four-dimensional Chern--Simons term is
automatically generated by the generalized reduction, thus preserving
the consistency of the theory.

The Scherk--Schwarz reduction is performed by imposing generalized 
periodicity conditions on the fields:
\beq
\Phi(x,y+2\pi r)= U \ \Phi(x,y) \, , \nn
\eeq
where $U$ is a constant matrix corresponding to a symmetry of the
five-dimensional theory. For the moment we consider only the case $U
\in USp(4)$. The case of a non-compact twist $U \in SO(1,1)$ will be
discussed separately at the end of this section: as we will see, in
such a case a positive-definite scalar potential without critical
points is generated, and the theory has no $\De4$ maximally symmetric
vacuum.

Since $USp(4)$ has rank two, we can parametrize the twist $U$ as
\beq
U=\exp{ \l[ i \l( \alpha_1 \frac{Y}{2}
+ \alpha_2 \frac{T_3}{2} \r) \r]} \, ,
\eeq
where $Y$ and $T_3$ are two representative generators in the Cartan
subalgebra of $USp(4)$, whose explicit representation is given in
appendix~\ref{app:conventions}, and $\alpha_{1,2} \in {\bf R}$.
Following the standard procedure, we reparametrize the twisted fields
in terms of periodic ones:
\be
\Phi(x,y) \equiv U(y) \ \widetilde{\Phi}(x,y) \, ,
\qquad
\widetilde \Phi(x,y) = \widetilde \Phi(x,y+ 2\pi r) \, ,	
\ee
\be
U(y) 
= 
\exp{ \l[ \frac{i \, y}{2\pi r} \l( \alpha_1 \frac{Y}{2}
+ \alpha_2 \frac{T_3}{2} \r) \r]}
=
\exp{\l[ i y \l( m_1 \frac{Y+T_3}{2}+m_2 \frac{Y-T_3}{2}
\r) \r]}
\equiv
e^{\dd i y M} 
\, .
\label{eq:twist}
\ee
For simplicity, tildes will be dropped from now on, and fields will
always be understood to be periodic.

With respect to the standard reduction, the generalized one produces
extra terms in the $\De4$ effective Lagrangian, proportional to
\beq
U^{-1}(y) \, \partial_y \, U(y) = i \ M \, .
\eeq
In our conventions, the explicit representations of the mass matrix
$M$, acting respectively on the $\bf 4$ and the $\bf 5$ of $USp(4)$,
are:
\bea
M_4 &=& {\rm diag}\l ( m_1 \, \sigma_3\,,m_2\,\sigma_3 \r) \, , \\
M_5 &=& {\rm diag}\l [ (m_1+m_2)\, \sigma_2\,,(m_1-m_2)\,\sigma_2
\,,0 \r] \, .
\eea
Some of the extra contributions to the $\De4$ Lagrangian correspond to
fermion mass terms. The remaining ones can be consistently organized
to describe the `gauging' of the $\Ne4$ theory: the upgrade of a
subgroup of the global $U$-duality group of the $\De4$ theory to a
local invariance.  The gauge group, however, is not the direct product
of simple and abelian factors. It can be identified with the
semidirect product of the $U(1) \subset USp(4)$ associated with the
twist $U$, and four translations: $U(1) \ \circledS \ {\cal T}^4$. The
associated Lie algebra is defined by the commutation relations:
\bea 
& \Bigl[ X_{\widehat{i}}, X_7 \Bigr] =
f^{\widehat{j}}_{\ \widehat{i} \,  7} 
X_{\widehat{j}} \, , 
\qquad 
\Bigl[ X_{\widehat{i}}, X_{\widehat{j}} \Bigr] = 0 \, ,
& \nn \\ & f^{\widehat{j}}_{\
\widehat{i} \, 7}= i M^{\widehat{j}}_{\ \ \widehat{i}} \, , &
\label{eq:gaugalg}
\eea
where $\widehat{i}=1\dots6$, $M^{\widehat{k}}_{\ \ \widehat{l}}={\rm
diag} \, [(M_5)^k_{\ \ l},0]$, $X_7$ is the twist generator (with vector
potential $A_\mu$), and the $X_i$ are the generators of $\bf 5^+$
(with vector potentials $B_\mu^i$). Notice that only four independent
$X_i$ enter non-trivially in the Lie algebra, as the fifth one
commutes with the twist. Moreover, the generator $X_6$ of $\bf 1^+$
always remains ungauged, since it belongs to the $SU(1,1)$ sector,
which does not carry charges with respect to $USp(4)$. In the gauged theory,
according to the algebra of \eq{eq:gaugalg}, covariant
derivatives replace the ordinary derivatives of the ungauged theory,
and covariant field strengths are modified accordingly:
\bea
\widehat{D}_\mu \varphi^i &=& \partial_\mu \varphi^i - i (M_5)^i_{\ j}
\ (B_\mu^j + \varphi^j \ A_\mu ) \, , \nn \\
\widehat{B}_{\mu \nu}^i &=& ( \partial_\mu B_\nu^i - \partial_\nu
B_\mu^i) - i (M_5)^i_{\ j} \ (A_\mu B_\nu^j - B_\mu^j \ A_\nu ) \, ,
\nn \\ \widehat{D}_\mu \chi_a &=& \partial_\mu \chi_a - i A_\mu \
(M_4)_a^{\ b} \ \chi_b \, , \nn \\ 
\widehat{D}_\mu \eta_{\nu \ a}
&=& \partial_\mu \eta_{\nu \ a} - i \ A_\mu \ (M_4)_a^{\ b} \
\eta_{\nu \ b} \nn \, , \\
\widehat{D}_\mu \psi_{y \ a}
&=& \partial_\mu \psi_{y \ a} - i \ A_\mu \ (M_4)_a^{\ b} \
\psi_{y \ b} \, . \label{eq:Dscal} 
\eea
Consequently, the transformation laws under the gauged symmetry,
with local parameters $\Xi^I$ ($I=1,\ldots,7$), become:
\bea
\delta \varphi^i &=& i (M_5)^i_{\; j} \, (\Xi^j + \varphi^j \ \Xi^7 )
\, , \nn \\ \delta B_\mu^i &=& \partial_\mu \Xi^i + i (M_5)^i_{\; j} \
( \Xi^7 \ B_\mu^j -\Xi^j \ A_\mu ) \, .
\eea
Under the gauged symmetry, the $\theta_{IJ}$ matrix of
\eq{eq:deftheta} transforms non linearly. This would require
the addition of an extra Chern--Simons term in order to guarantee 
the gauge invariance of the $\De4$ theory \cite{afl}, namely:
\beq
- {2 \over 3} i \ d_{\widehat{\imath}\, \widehat{\jmath} \,
\widehat{k}}\, M^{\widehat{k}}_{\ \ \widehat{l}} \,
\varepsilon^{\mu \nu \rho \sigma}  B_\mu^{\widehat{\imath}}
B_\nu^{\widehat{l}} B_{\rho \sigma}^{\widehat{\jmath}} \, ,
\eeq
where $d_{\widehat{\imath}\, \widehat{\jmath} \, \widehat{k}}$ is a
symmetric $SO(5)$-invariant tensor, normalized to:
\beq
d_{ij6}=d_{i6j}=d_{6ij}= - {1 \over 4} \delta_{ij}\, . \nn
\eeq
As argued in \cite{gau8a}, and verified explicitly in the present
case, this term is automatically produced by the generalized
dimensional reduction.

The detailed expression of the gauged Lagrangian can be found in
appendix~\ref{app:4dlagss}. We display here only its bosonic part:
\bea
e_4^{-1} \cL_{\rm bos}^{SS} &=& 
-R_4- {1 \over 2} { \partial_\mu t \partial^\mu t 
+ \partial_\mu \tau \partial^\mu \tau \over t^2}
- { \partial_\mu \varphi_0 \partial^\mu \varphi_0
+\widehat{D}_\mu \varphi_i \widehat{D}^\mu \varphi^i 
\over \varphi_0^2} 
\nn \\
&&
-\frac14  \, g_{IJ} \, \widehat{F}^I_{\mu\nu} 
\widehat{F}^{J \ \mu\nu} -\frac18 \, e_4^{-1} \theta_{IJ} \, 
\varepsilon^{\mu\nu\rho\sigma} \widehat{F}^I_{\mu\nu} 
\widehat{F}^{J}_{\rho\sigma} 
\nn \\
&&
- {2 \over 3} i \ d_{\widehat{\imath}\, \widehat{\jmath} \,
\widehat{k}}\, M^{\widehat{k}}_{\ \ \widehat{l}} \,
\varepsilon^{\mu \nu \rho \sigma}  B_\mu^{\widehat{\imath}}
B_\nu^{\widehat{l}} B_{\rho \sigma}^{\widehat{\jmath}} \, .
\eea
For generic values of $m_1$ and $m_2$, the gauging of the four
translations allows us to shift away four spin-0 fields: they are
absorbed by the corresponding vectors, which acquire a mass matrix
proportional to $M_5$. For $|m_1| = |m_2|$, only two spin-0 fields are
absorbed by the vectors gauging the Euclidean group on the plane,
$U(1) \ \circledS \ {\cal T}^2$.

In the fermionic sector, besides the covariantization of derivatives
and field strengths with respect to the gauged group, mass terms
appear, proportional to $M_4$, and a super-Higgs effect takes place,
with the ${\psi_y}_a$ playing the role of goldstinos. This can be
deduced, for instance, by looking at the inhomogeneous terms in the
supersymmetry transformations of \eq{eq:susytransf4d}.

Depending on the specific choice of the mass parameters $m_1$ and
$m_2$ appearing in the twist matrix, it is possible to break
supersymmetry partially ($\Ne2$) or completely ($\Ne0$). We
summarize the spectrum for the different cases in our short-hand
notation:
\bea
m_1 \ne m_2 = 0 &:& [1,2,1,0,0]^{{\cal N}=2}_{m=0}+ \l\{ 2 \times
[0,1,2,1,0]^{{\cal N}=2}_{m \ne 0} \r\}+ 2 \times [0,0,1,2,2]^{{\cal
    N}=2}_{m=0} \, , \nn \\ |m_1|\neq |m_2| ,m_1 m_2\neq 0 &:&
[1_{m=0}, 4_{m \ne 0}, 4_{m \ne 0} + 3_{m=0}, 4_{m \ne
    0},4_{m=0}]^{{\cal N}=0} \,, \nn \\ |m_1|= |m_2|\ne 0 &:&
[1_{m=0}, 4_{m \ne 0}, 2_{m \ne 0} + 5_{m=0}, 4_{m \ne
    0},6_{m=0}]^{{\cal N}=0} \,. \nn
\eea
In the partially broken case, beside the gravitational multiplet and
two vector multiplets of $\Ne2$, we can recognize a massive $\Ne2$
BPS short multiplet of spin $3/2$ \cite{ferrara:superhiggs}.

The spectrum of the theory can be easily extracted from the Lagrangian
in \eq{eq:lag4dss}. For the bosonic sector, it is useful to observe
that the kinetic terms for the vector fields are diagonal in the basis
of the $V_\mu^{\widehat i} = B_\mu^{\widehat i} +\varphi^{\widehat i}
A_\mu$:
\be
 - \frac14 \ t \ \overline{V}_{\mu\nu}^{i} \overline{V}^{\mu\nu}_{i} 
- \frac14 \frac{\varphi_0^2}{2\,t} \, \overline{v}_{\mu \nu}
 \overline{v}^{\mu \nu} - \frac14 \frac{\varphi_0^2\,t}{2}\,
 A_{\mu\nu} A^{\mu\nu}\, ,
\ee
through covariantization with respect to the vector metric $g_{IJ}$:
\bea
\overline{V}_{\mu\nu}^{\widehat i} &\equiv& 2{\cal D}_{[\mu}
V_{\nu]}^{\widehat i} \, , \nn \\ {\cal D}_\mu V_\nu^{\widehat i} &\equiv&
\partial_\mu V_\nu^{\widehat i} - (\partial_\mu \varphi^{\widehat i})
\ A_\nu + i (M_5)^{\widehat i}_{\; \widehat{j}} \ V_\mu^{\widehat j}
A_\nu = \partial_\mu V_\nu^{\widehat i}- ({\cal D}_\mu
\varphi^{\widehat i}) \ A_\nu \, . 
\eea
The spectrum is twice degenerate and reads:
\bea
{\rm spin \; 3/2} \ &(\eta_\mu^{1\dots4}) &: \qquad { 2 \over
t\varphi_0^2 } m_{1,2}^2 \, , \nn \\ {\rm spin \; 1} \ &(V_\mu^{1\dots4})
&: \qquad \frac{2}{t \varphi_0^2} (m_1\pm m_2)^2\, , \nn \\ {\rm
spin \; 1/2} \ &(\chi^{1\dots4})&: \qquad { 2 \over t\varphi_0^2 }
m_{1,2}^2 \, , \nn
\eea
{}from which we can check the $\Ne4$ mass sum rule $Str \ {\cal
M}^2=0$.

\subsection{Non-compact $SO(1,1)$ twist}

If non-compact generators are used to perform the Scherk-Schwarz
reduction, non-flat gaugings are generated \cite{ss}, with a
positive-definite four-dimensional potential without critical
points. This is the case if we use the non-compact $SO(1,1)$ in the
$U$-duality group for the twist, in analogy with \cite{ncgau}.

Given the SO(1,1) field transformations in eq.~(\ref{eq:so115d}), we
obtain the following relations between periodic (with tildes) and
non-periodic (without tildes) fields:
\be
V_M^i = e^{\Lambda\,y} \ \widetilde V_M^i \, ,
\quad
v_M = e^{-2 \, \Lambda \, y} \ \widetilde v_M \, ,
\quad
X = e^{\Lambda \, y} \ \widetilde X \, .
\ee
After the $SO(1,1)$ generalized dimensional reduction, and removing
the tildes as usual, we get the following $\De4$ bosonic Lagrangian:
\bea
e_4^{-1}\cL^{SO(1,1)}_{\rm bos}&=& 
-R_4 -\frac12 \frac{D_\mu t D^\mu t+D_\mu \tau D^\mu \tau}{t^2} 
-\frac{D_\mu \varphi_0 D^\mu \varphi_0+D_\mu \varphi_i 
D^\mu \varphi_i}{\varphi_0^2} \nn \\
&&-\frac14 g_{IJ}\widehat F_{\mu\nu}^{I} \widehat F^{J \ \mu\nu}
-\frac18 e_4^{-1} \theta_{IJ} \varepsilon^{\mu\nu\rho\sigma} 
\widehat F_{\mu\nu}^{I} \widehat F_{\rho\sigma}^{J} 
\nn \\
&&-\frac{1}{6} e_4^{-1} \Lambda \, {\cal C}_{IJK} \, 
\varepsilon^{\mu\nu\rho\sigma} B_{\mu}^{I} B_{\nu}^{J} 
\widehat F_{\rho\sigma}^{K} 
- \frac{6 \Lambda^2}{t \varphi_0^2} \, ,
\label{eq:so11lagbos}
\eea
where the coefficients ${\cal C}_{IJK}$ are fully symmetrized. The 
non-vanishing independent ones are identified by
\be
{\cal C}_{i67} = \varphi_i \, , 
\qquad 
{\cal C}_{ij6}=-\delta_{ij} \, , 
\qquad 
{\cal C}_{677}=\varphi_i^2 \, .
\ee
The scalar covariant derivatives are
\bea
&& D_\mu \varphi_0 = \l (\partial_\mu-\Lambda A_\mu\r)\varphi_0 \,, 
\qquad D_\mu \varphi^i=\partial_\mu \varphi^i - \Lambda \l 
( B_\mu^i+ \varphi^i A_\mu  \r) \,, \nn \\
&&D_\mu t=\l (\partial_\mu+2\Lambda A_\mu\r)t \,, 
\qquad D_\mu \tau=\partial_\mu \tau +2\Lambda \l 
(b_\mu+\tau A_\mu\r) \, ,
\eea
and the covariant field strengths are
\bea
\widehat F^i_{\mu\nu} & = & B_{\mu\nu}^i-\Lambda \l 
(A_\mu B_\nu^i-B_\mu^i A_\nu\r) \, , \nn \\
\widehat F^6_{\mu\nu} & = & b_{\mu\nu}+2\Lambda \l 
(A_\mu b_\nu-b_\mu A_\nu\r)\,, \nn \\
\widehat F^7_{\mu\nu} & = & A_{\mu\nu}\, . 
\eea
{}From \eq{eq:so11lagbos}, we see that the $SO(1,1)$ twist produces a
positive-definite, non-vanishing scalar potential without critical
points, which does not admit maximally symmetric $\De4$ vacua.

In the fermionic Lagrangian, all the field strengths and the scalar
derivatives become covariant as described above. The fermions remain
neutral with respect to these gauge interactions, but acquire a
field-dependent mass term controlled by the scalar potential:
\beq -\frac12 \l (\frac{6 \Lambda^2}{t \varphi_0^2} \r)^{1/2} \bar
\eta_\mu^a \widehat \gamma \gamma^\mu \chi_a \, .
\eeq

The theory now has a non-abelian gauge group, which is the semidirect
product of the six translations ($B_\mu^i,b_\mu$) and the dilatation
($A_\mu$), namely $SO(1,1) \ \circledS \ {\cal T}^6$. All the seven
vectors acquire a field-dependent mass term proportional to the
potential, absorbing the corresponding seven spin-0 fields. The
surviving scalar, associated with the ungauged $SO(1,1)$, has a
runaway behaviour described by the potential of \eq{eq:so11lagbos}.

%%%%%%%%%%%%%%%%%%%%%%%%%%%%%%%%%%%%%%%%%%%%%%%%%%%%%%%%%%%%%%%%%%%%
%%                          SECTION                               %%
%%%%%%%%%%%%%%%%%%%%%%%%%%%%%%%%%%%%%%%%%%%%%%%%%%%%%%%%%%%%%%%%%%%%

\section{ \boldmath{$\Ne4$}, \boldmath{${\it D} \ge 6$} reductions}

The discussion given in section~\ref{sec:SSred} can be extended to higher
dimensions. Starting from a $\Ne4$ pure ungauged supergravity in $D$
dimensions, by generalized dimensional reduction we can obtain a
minimal~\footnote{As we will see below, in six dimensions there exist
two different $\Ne4$ theories, and only one of them gives the
`minimal' $\De5$ no-scale model.} no-scale model in $D-1$
dimensions. However, this strategy works only for $D \le 8$, where
there is a non-trivial $R$-symmetry for the twist. As an example, in
this section we will show that from the generalized reduction of an
ungauged $\Ne4$, $\De6$ pure supergravity we can obtain some new
$\Ne4$, $\De5$ gauged supergravities, spontaneously broken to $\Ne2$
or $\Ne0$.  These theories were not considered in a previous
classification \cite{dagata} that concentrated on semisimple and
abelian gaugings.  Analogous results have recently been found
\cite{afl5} for the $\Ne8$ and $\Ne2$ cases, and the present
discussion will complete that analysis. Since there exist two
inequivalent $\Ne4$, $\De6$ supergravities, the non-chiral one, or
(2,2) \cite{giani}, and the chiral one, or (4,0) \cite{townsend}, we
now discuss them in turn. We then conclude the section with a brief
discussion of the remaining cases, $D=7 \to 6$ and $D = 8 \to 7$,
which complete the analysis of such gaugings for minimal $\Ne4$
theories.

\subsection{Reduction of $\De6$ (2,2) supergravity}

The $R$-symmetry of the $\De6$ (2,2) non-chiral supergravity is
$USp(2) \times USp(2)$. The scalar manifold of the generic theory with
$n$ vectors is:
\beq
{SO(4,n)\over SO(n)\times SO(4)}\times SO(1,1) \, ,
\eeq
which reduces to $SO(1,1)$ in the pure $n=0$ case.  The $U$-duality
group is then $SO(1,1) \times SO(4)$ [with $SO(4) \sim SO(3) \times
SO(3) \sim USp(2) \times USp(2)$]. The two-parameter twist can be
constructed with two generators of $SO(4)$, one for each $USp(2)$
subgroup. The (2,2) gravitational multiplet is:
\beq
 _{\De6}[1,(2,1)+(1,2),(2,2)+(1,1),(2,1)+
(1,2),1]^{{\cal N}=4}_{m=0} \, ,
\eeq
where we have made explicit the representations under $USp(2) \times
USp(2)$, and the spin-1 (1,1) entry is actually an antisymmetric
2-form (or equivalently one self-dual and one anti-self-dual tensor).

The 2-form is inert under the $R$-symmetry, and after reduction to
$\De5$ it produces one vector and one 2-form, which in $\De5$ can be
dualized to another vector. The remaining four vector fields are all
charged and participate in the gauging of the $U(1) \ \circledS \
{\cal T}^4$ group, with the graviphoton $e_M^{\ \ \widehat{6}}$
associated with the $U(1)$ factor. The embedding of the gauged group
can be easily understood by looking at the decomposition of the
$\Ne4$, $\De5$ $U$-duality algebra:
\beq
SO(1,1)\times SO(5,1)\rightarrow 
SO(1,1)+SO(1,1)+SO(4)+{\bf 4}^+ +{\bf 4}^- \, .
\eeq
We thus get the following $\De5$ theories:
\bea
m_1 \ne m_2=0 &:& [1,2,1,0,0]^{{\cal N}=2}_{m=0}+ \l\{ 2 \times
[0,1,2,1,0]^{{\cal N}=2}_{m \ne 0} \r\}+ 2 \times [0,0,1,2,1]^{{\cal
    N}=2}_{m=0} \, , \nn \\ |m_1|\neq |m_2| ,m_1 m_2\neq 0 &:&
[1_{m=0}, 4_{m \ne 0}, 4_{m \ne 0} + 3_{m=0}, 4_{m \ne
    0},2_{m=0}]^{{\cal N}=0} \,, \nn \\ |m_1|=|m_2|\neq 0 &:&
[1_{m=0}, 4_{m \ne 0}, 2_{m \ne 0} + 5_{m=0}, 4_{m \ne
    0},4_{m=0}]^{{\cal N}=0} \, ,
\eea
where the $\Ne2$ and $\Ne0$ cases are obtained by appropriate choices
of the twist parameters in the two $USp(2)$. The masses of fermions
and vectors have the same dependence on the twist parameters as in the
$\De4$ case.

\subsection{Reduction of $\De6$ (4,0) supergravity}
Pure $\Ne4$, $\De6$ chiral supergravity is anomalous, and extra
multiplets have to be added to obtain a consistent theory. The choice
is unique \cite{townsend} and consists in adding 21 antisymmetric
tensor multiplets. This produces a theory with the following field
content:
\beq
_{\De6}[1,4,5^-,0,0]_{m=0}^{\Ne4} \quad + \quad 21 \times \
_{\De6}[0,0,1^+,4,5]_{m=0}^{\Ne4} \, .
\eeq
The spin-1 entries $5^-$ and $1^+$ are (anti) self-dual tensors,
transforming in the {\bf 5} and in the {\bf 1} of $USp(4)$,
respectively. The theory then has 105 scalars, parametrizing the
manifold:
\beq
\frac{SO(5,21)}{USp(4)\times SO(21)} \, .
\eeq
We could perform a Scherk--Schwarz twist with a generator in the
maximal compact subgroup of the $U$-duality group, namely $USp(4)
\times SO(21)$. However, we consider only a twist in the $USp(4)$
factor, since a twist in $SO(21)$ does not break supersymmetry. Notice
also that no vectors are present in this $\De6$ theory, and that the
only spin-1 fields charged under $USp(4)$ are anti-self-dual tensors.
This means that in the present example there are no shift symmetries
to be gauged. Indeed, the only gauged symmetry can be a $U(1)$
subgroup of $USp(4)$, via the vector potential $e_M^{\ \ \widehat
6}$. The generalized reduction can then be performed as in the $\De5$
case, by choosing the twist of eq.~(\ref{eq:twist}). In the $\De5$
reduced theory, the fermions in the {\bf 4} of $USp(4)$ acquire mass
as in the $\De4$ case. Four of the $5^-$ tensors transform into two
complex $\De5$ tensors ($2^{\bf c}$) with masses $m_1\pm m_2$. The
other 22 tensors remain massless and neutral, so that they can be
dualized to $\De5$ vectors. Finally, the scalars acquire a potential,
which vanishes at its minimum. The potential has 21 flat directions,
corresponding to 21 massless scalars. The other 42+42 scalars acquire
instead a mass $m_1\pm m_2$.

In terms of $\De5$ multiplets, the spectrum can be summarized 
as follows:
\bea
m_1\ne m_2=0 &:& [1,2,1,0,0]^{{\cal N}=2}_{m=0}+ [0,2,2^{\bf
c},0,0]^{{\cal N}=2}_{m \ne 0} + 22 \times [0,0,1,2,1]^{{\cal
N}=2}_{m=0} \nn \\ && + 21 \times [0,0,0,2,4]^{{\cal N}=2}_{m\neq0}\,
, \nn \\ |m_1|\neq |m_2| ,m_1 m_2\neq 0 &:& [1_{m=0}, 4_{m \ne 0},
2^{\bf c}_{m\ne0} + 23_{m=0} , 84_{m \ne 0},84_{m \ne
0}+21_{m=0}]^{{\cal N}=0} \,, \nn \\ |m_1|=|m_2|\ne 0 &:& [1_{m=0},
4_{m \ne 0}, 1^{\bf c}_{m \ne 0} + 25_{m=0}, 84_{m \ne
0},42_{m\ne0}+63_{m=0}]^{{\cal N}=0} \, ,
\eea
where the massive fields are also charged with respect to the local
$U(1) \subset USp(4)$.

We notice finally that, starting from the (anomalous) pure $\De6$
(4,0) supergravity, or performing a consistent truncation of the
reduced theory, it is possible to recover the flat pure $\De5$ gauged
theory of \cite{romans}. This happens because the reduction of the
pure (4,0) theory does not produce extra matter in $\De5$.  In
particular, we can obtain the flat case of \cite{romans} by setting
$m_1=m_2$. However, we can also choose $m_1 \ne m_2$, which leads to
two massive complex anti-self-dual tensors. In particular, the case
$m_2=0$ gives partial breaking of supersymmetry. All this generalizes
the minimal $\Ne4$, $\De5$ no-scale model given in \cite{romans}.

\subsection{Reductions from $\De7$ and $\De8$}
We complete the higher-dimensional case with a brief discussion of
$\De7$ and $\De8$. In pure $\Ne4$, $\De7$ supergravity, we have the global
$USp(2) \times SO(1,1)$ group, and the gravitational multiplet is:
\beq
_{\De7}[1,2,1_2+3,2,1]_{m=0}^{\Ne4} \, ,
\eeq
where we made explicit the $USp(2)$ irreducible representations. The
spinors are symplectic Majorana in $\De7$, and the spin-1 singlet
($1_2$) is a 2-form. Standard reduction gives a non-chiral $USp(2)$,
$\Ne4$, $\De6$ supergravity~\footnote{This $USp(2)$ is the diagonal
subgroup of the full $\De6$ $R$-symmetry group, $USp(2)\times USp(2)$.}
coupled to one vector multiplet:
\beq
_{\De6}[1,2+2,1_2+1+3,2+2,1]_{m=0}^{\Ne4}
\ + \ _{\De6}[0,0,1,2+2,1+3]_{m=0}^{\Ne4} \, ,
\eeq
with scalar manifold
\beq
\frac{SO(4,1)}{SO(4)} \times SO(1,1) \, .
\eeq
The algebra of the duality group decomposes into
\beq
SO(1,1) + SO(1,1) + USp(2) + \bf{3}^+ + \bf{3}^- \, .
\eeq
After a twist in $USp(2)$, we end up with:
\beq
_{\De6}[1_{m=0},4_{m\ne0},(1_2)_{m=0}
+ 2_{m=0} + 2_{m\ne0},4_{m\ne0},3_{m=0}]^{\Ne0} \, ,
\eeq
the fermions getting a mass $m$ proportional to the twist, and
two vectors getting a mass $2m$, via the gauging of the semidirect
product $U(1) \ \circledS \ {\cal T}^2$.

Finally, the $\De8$ theory is globally invariant only under 
$SO(2) \times SO(1,1)$. With respect to $SO(2)$, the gravitational 
multiplet can be decomposed as
\beq
_{\De8}[1,1^+,1_2+2,1^-,1]_{m=0}^{\Ne4} \, ,
\eeq
where $(1^\pm)$ means a complex Weyl spinor charged under $U(1) \sim
SO(2)$.  As in the previous case, the reduced theory is coupled to one
vector multiplet:
\beq
_{\De7}[1,2,1_2+1+2,2,1]_{m=0}^{\Ne4} \ + 
\ _{\De7}[0,0,1,2,1+2]_{m=0}^{\Ne4} \, .
\eeq
The scalar manifold is
\beq
\frac{SO(3,1)}{USp(2)}\times SO(1,1) \, ,
\eeq
and the algebra of the duality group decomposes into 
\beq 
SO(1,1) + SO(1,1) + U(1) + \bf{2}^+ + \bf{2}^- \, .
\eeq
Again, a $U(1)$ twist leads to the gauging of
the subgroup $U(1) \ \circledS \ {\cal T}^2$, with matter content
\beq
_{\De7}[1_{m=0},2_{m\ne0},(1_2)_{m=0} +
2_{m=0}+2_{m\ne0},4_{m\ne0},2_{m=0}]^{\Ne0} \, ,
\eeq
fermions of mass $m$ and vectors of mass $2m$.

%%%%%%%%%%%%%%%%%%%%%%%%%%%%%%%%%%%%%%%%%%%%%%%%%%%%%%%%%%%%%%%%%%%%
%%                          SECTION                               %%
%%%%%%%%%%%%%%%%%%%%%%%%%%%%%%%%%%%%%%%%%%%%%%%%%%%%%%%%%%%%%%%%%%%%

\section{\boldmath{${\cal Z}_2$} orbifold}

We have seen that generalized dimensional reduction can give either
partial or complete breaking of supersymmetry, but the number of
residual supersymmetries is always even, as the mechanism cannot
generate chirality. However, with the additional help of an orbifold
projection, it is also possible to obtain a reduced theory with an odd
number of unbroken supersymmetries. It is then interesting to study
how the results obtained in section~\ref{sec:SSred} can be modified by
orbifold projections. With only one internal dimension, it is not
restrictive to consider the \Zd \ orbifold associated with the parity
$y \to - y$, which leaves the classical $\De5$ action invariant. The
corresponding action on the fields can be written as:
\be 
\Phi(x^\mu,-y) = {\cal Z}_2 \ \Phi(x^\mu,y) \, ,
\ee
where for consistency \Zd \ must square to 1. In the absence of a
twist, a consistent assignment of the \Zd \ parities to the fields is
the following:
\bea
e_\mu^\alpha &:& + \nn \\
\rho\,,\phi\,,v_5 &:& + \nn \\
A_\mu\,,v_\mu &:& - \nn \\
V_\mu^i &:& (+,+,-,-,-) \nn \\
V_5^i &:& (-,-,+,+,+) \nn \\
{\psi_\mu}_a &:& (+,-,+,-) \nn \\
{\psi_y}_a\,,\chi_a &:& (-,+,-,+) \, . 
\eea
In our conventions, the above assignment corresponds to the following
\Zd \ representation:
\beq  \label{eq:z2}
{\psi_\mu}_a (-y)=\widehat \gamma \, Y_a^{\ b} {\psi_\mu}_b (y) \, ,
\eeq
where $Y$ is the $U(1) \subset USp(4)$ generator given in appendix~A.
All other parity assignments follow from \eq{eq:z2}, and the other 
admissible choices for \Zd \ are physically equivalent.

The standard reduction on the orbifold $S^1/{\cal Z}_2$ produces an
unbroken $\Ne2$, $\De4$ theory, with one gravitational multiplet
[$e_\mu^\alpha, \psi_\mu^{1,3}, V^1_\mu$], one vector multiplet
[$V^2_\mu, \psi_y^{2,4}, V_5^{3,4}$] and one hypermultiplet
[$\chi^{2,4}, (\rho, \phi, v_5, V_5^5)$]:
\be
[1,2,1,0,0]^{{\cal N}=2}_{m=0} \ + \ 
[0,0,1,2,2]^{{\cal N}=2}_{m=0} \ + \
[0,0,0,2,4]^{{\cal N}=2}_{m=0} \, .
\ee

We are now ready to discuss what happens if the \Zd \ orbifold
projection and generalized dimensional reduction are combined.  For a
consistent space-time interpretation, the orbifold symmetry \Zd \ and
the twist $U$ must obey the relation:
\beq
{\cal Z}_2 \, U \, {\cal Z}_2 \, U = 1 \, .
\label{eq:z2Ucons}
\eeq
If we express the twist as $U=\exp (i \ T)$, we can identify two ways
of satisfying eq.~(\ref{eq:z2Ucons}): either $[T,{\cal Z}_2]=0$ or
$\{T,{\cal Z}_2\}=0$. In the first case, eq.~(\ref{eq:z2Ucons})
reduces to $U^2=1$, which discretizes the possible values of the twist
$T$. In the second case, instead, eq.~(\ref{eq:z2Ucons}) is satisfied
for every value of the twist.

In both cases the gauging identified in section~\ref{sec:SSred} is
nullified by the orbifold, since the projection removes the zero mode
of $A_\mu$, which gauges $USp(4)$ and is crucial for the non-abelian
character of the algebra in eq.~(\ref{eq:gaugalg}). However, there is
still room for a smaller abelian group to be gauged. We therefore
analyse the two cases in more detail.

To satisfy $[T,{\cal Z}_2]=0$, we must search for generators of
$USp(4)$ that commute with $Y$, but the most general twist with these
properties is the one already analysed in section~\ref{sec:SSred}. To
satisfy also $U^2=1$, we must choose $\alpha_1 \pm \alpha_2 = 0$
(mod.~$2 \pi$), or, equivalently, $m_{1,2} = (0, \pm 1)/(2r)$. If we
do so, also the twist acts like a parity, in particular $U={\cal
Z}_2\cdot{\cal Z}_2^\prime$, where ${\cal Z}_2$ is the reflection with
respect to $y=0$ and ${\cal Z}_2^\prime$ the one with respect to
$y=\pi$. This is equivalent to imposing a ${\cal Z}_2\times{\cal
Z}_2^\prime$ projection on a circle of radius $2r$. Choosing for
instance $m_1=0$, $m_2=1/(2r)$, the $({\cal Z}_2\,,{\cal Z}_2^\prime)$
parity assignments will be:
\beq
\barr{cccccc} 
e_\mu^\alpha\,,\rho\,,\phi\,,v_5 & {\psi_\mu}_a & {\psi_y}_a \,,\chi_A
& A_\mu\,,v_\mu & V_\mu^i & V_5^i \\ (+,+)& \l
(\barr{c}+,+\\-,-\\+,-\\-,+ \earr\r) & \l (\barr{c}-,-\\+,+\\-,+\\+,-
\earr\r) & (-,-) & \l (\barr{c}+,-\\+,-\\-,+\\-,+\\-,- \earr\r) & \l
(\barr{c}-,+\\-,+\\+,-\\+,-\\+,+ \earr\r)
\earr
\, .
\eeq
This means that the reduced theory is an ${\cal N}=1$, $\De4$
unbroken supergravity with one gravitational multiplet [$e_\mu^\alpha,
{\psi_\mu}_1$] and two chiral multiplets [$({\psi_y}_2, \chi_2),
(\rho, \phi, v_5, V_5^5)$]:
\be
[1,1,0,0,0]^{{\cal N}=1}_{m=0} \ \ + \ \ 2 
\ \times[0,0,0,1,2]^{{\cal N}=1}_{m=0} \, .
\ee

We can finally move to the more interesting case $\{T,{\cal
Z}_2\}=0$. Six generators of $USp(4)$ have this property, namely those
in the coset $USp(4) / [SU(2) \times U(1)]$. We can easily find two of
these generators that commute with each other (and obviously
anticommute with $Y$). Without loss of generality we choose:
\bea
T_{31}&=&-i \, \Gamma_{31}=1\otimes\sigma_2 \nn \, , \\
T_{24}&=&-i \, \Gamma_{24}=\sigma_3\otimes\sigma_2 \, .
\eea
Our twist then reads:
\beq
U=\exp\l [i \l(\alpha_1 \frac{T_{31}}{2}
+ \alpha_2 \frac{T_{24}}{2}\r) \r ]
\, .
\eeq
The action of the twist gives the same results as those in
section~\ref{sec:SSred}, but with different matrices $M_{4,5}$. In
particular, we have now:
\bea
M_4&=&{\rm diag} \Bigl ( m_1\, \sigma_2 \,, m_2\, \sigma_2 \Bigr) \,,
\nn \\ M_5&=&{\rm diag} \Bigl ( -(m_1+m_2) {(\sigma_2)}_{13} \,,
(m_1-m_2) {(\sigma_2)}_{24} \,, (0)_5 \Bigr) \,, \label{eq:M45orbi}
\eea
where $(\sigma_i)_{ij}$ means a $\sigma_i$ matrix taken in the
subspace $(_{ij})$.

The \Zd \ orbifold projection removes some fields from the reduced
theory, in particular the field $A_\mu$, responsible for the gauging
described in section~\ref{sec:SSred}. However, only half of the
gauged translations are removed, while the others survive. Observe
that the orbifold preserves only the vectors $V^{1,2}_\mu$ and
the scalars $V^{3,4,5}_5$. However, now the twist connects $1 \to 3$
and $2 \to 4$, still allowing the surviving scalars to be gauged. This
can be explicitly checked in the covariant derivative for the scalars,
eq.~(\ref{eq:Dscal}):
\beq
\widehat{D}_\mu \varphi^{3,4} = \partial_\mu \varphi^{3,4} - i
(M_5)^{3,4}_{\quad {1,2}} \ V_\mu^{1,2} \, .  
\eeq

The mass matrices in eq.~(\ref{eq:M45orbi}) are non-diagonal, but
their squares are:
\bea
M^2_4&=&{\rm diag} \l ( m_1^2  \,, m_1^2  \,, m_2^2 \,, 
m_2^2 \r) \,, \nn \\
M^2_5&=&{\rm diag} \l ( (m_1+m_2)^2  \,, 
(m_1-m_2)^2 \,, (m_1+m_2)^2 \,, (m_1-m_2)^2 \,, 0 \r) \, .
\eea
The \Zd \ orbifold projection removes the second and the fourth entry
in $M_4$, as well as the last three entries in $M_5$. The resulting 
$\Ne2$ theory is spontaneously broken either to $\Ne1$ or to $\Ne0$, 
with two independent mass parameters, and inherits the $\Ne4$ mass 
sum rule $Str \, {\cal M}^2 = 0$.

The spectrum of the $\Ne4$, $\De5$ theory, after the orbifold and the
partial ($m_1\ne0,m_2=0$) or total ($m_1 m_2 \ne 0$)
Scherk--Schwarz breaking, reads:
\bea
&\Ne4, \ \De5& \nn \\ &\downarrow& \hspace{-4.0cm} {\cal Z}_2 \nn \\ \nn \\
&[1,2,1,0,0]^{{\cal N}=2}_{m=0}+[0,0,1,2,2]^{{\cal
N}=2}_{m=0}+[0,0,0,2,4]^{{\cal N}=2}_{m=0}& \nn \\ \nn \\ &\downarrow&
\hspace{-4.0cm} SS \ (m_1 \ne 0, m_2=0) \nn \\ \nn \\ &[1,1,0,0,0]^{{\cal
N}=1}_{m=0}+[0,1,2,1,0]^{{\cal N}=1}_{m\neq0}+2\times [0,0,0,1,2]^{{\cal
N}=1}_{m=0}& \nn \\ \nn \\ &\downarrow& \hspace{-4.0cm} SS \
(m_1 m_2\ne0) \nn \\ \nn \\
&[1_{m=0},2_{m\neq0},2_{m\neq0},2_{m\neq0},4_{m=0}]^{{\cal N}=0}& \nn
\eea
Notice that in the partially broken theory the massive multiplet is a
long spin-3/2 multiplet, because this is the only possibility with
${\cal N}=1$ (the short one, being BPS-charged, requires an even
number of supersymmetries).

This mechanism, already considered in \cite{alten}, allows us to break
an $\Ne2$ supergravity, via generalized reduction, with two
independent parameters, starting from the \Zd \ orbifold of an ${\cal
N}=4$, $\De5$ theory. The corresponding $\De4$ effective theory is the
minimal $\Ne2$ no-scale model with partial breaking of \cite{fgp}.

If we start from $\Ne8$, $\De5$ supergravity, we can get a $\Ne4$,
$\De4$ spontaneously broken supergravity with four independent mass
parameters, recovering the results of \cite{gau4b} for a Type IIB
supergravity compactified on a $T^6/{\cal Z}_2$ orientifold with 
fluxes.

%%%%%%%%%%%%%%%%%%%%%%%%%%%%%%%%%%%%%%%%%%%%%%%%%%%%%%%%%%%%%%%%%%%%
%%                          SECTION                               %%
%%%%%%%%%%%%%%%%%%%%%%%%%%%%%%%%%%%%%%%%%%%%%%%%%%%%%%%%%%%%%%%%%%%%

\section{Conclusions and outlook}

In this paper we constructed the minimal $\Ne4$ no-scale model in four
dimensions, by generalized dimensional reduction of pure $\Ne4$,
$\De5$ supergravity. We explicitly derived Lagrangian and
transformation laws of the $\De4$ effective theory, a $\Ne4$
supergravity where a flat, non-semisimple group is gauged, and half or
all of the supersymmetries are spontaneously broken. We found
that the Scherk--Schwarz reduction automatically generates the extra
Chern--Simons term that must be added to the $\De4$ Lagrangian to
ensure gauge invariance.  We also studied how this procedure extends
to non-compact twists and to higher dimensions, and found new $\Ne4$
gauged supergravities not included in earlier classifications.
Finally, we discussed the consistent orbifold projections of the
theory, recovering the minimal partially broken $\Ne2$ no-scale model
of refs.~\cite{fgp,alten}.

An interesting extension of the present work would consist in
performing a similar investigation with pure but gauged $\Ne4$, $\De5$
supergravity \cite{awatow, romans} as the starting point. This may
lead to a better understanding of spontaneous supersymmetry breaking
and boundary actions for warped compactifications, in the context of
the Randall--Sundrum model \cite{rs} with an underlying $\Ne4$
supersymmetry \cite{extrs}, explicitly broken to $\Ne2$ by the \Zd \
orbifold projection.

%
%%%%%%%%%%%%%%%%%%%%%%%%%%%%%%%%%%%%%%%%%%%%%%%%%%%%%%%%%%%%%%%%%%%%
%%                          SECTION                               %%
%%%%%%%%%%%%%%%%%%%%%%%%%%%%%%%%%%%%%%%%%%%%%%%%%%%%%%%%%%%%%%%%%%%%
%
\acknowledgments
We are grateful to S.~Ferrara for many illuminating discussions. We
also thank G.~Dall' Agata and J.-P.~Derendinger for discussions. This
work was partially supported by the European Programme
HPRN-CT-2000-00148 (Across the Energy Frontier).

%
%%%%%%%%%%%%%%%%%%%%%%%%%%%%%%%%%%%%%%%%%%%%%%%%%%%%%%%%%%%%%%%%%%%%
%%                          SECTION                               %%
%%%%%%%%%%%%%%%%%%%%%%%%%%%%%%%%%%%%%%%%%%%%%%%%%%%%%%%%%%%%%%%%%%%%
%
\appendix
\section{Conventions} 
\label{app:conventions}
We specify here our conventions on $USp(4)$ transformations and on
space-time spinors.
\subsection{$USp(4)$: group, algebra and representations}
The generic $USp(4)$ group element is defined as $4 \times 4$
complex matrix $U \equiv U_a^{\quad b}$ such that:
\beq 
U^\dagger \ U = 1 \, , 
\qquad \qquad
U^T \ \Omega \ U = \Omega \, ,
\label{usp4g}
\eeq
where $\Omega \equiv \Omega^{ab}$ is for us a real, antisymmetric
symplectic metric with upper indices:
\bea 
\Omega^{ab} \equiv \l(\barr{c c c c} 0 & 1 & 0 & 0 \\ 
-1 & 0 & 0 & 0 \\0 & 0 & 0 & 1 \\0 & 0 & -1 & 0 \earr \r) 
= \id_{2\times2} \otimes i\sigma^2 \, .
\label{smet} 
\eea 
The matrix $U$ acts from the left on column,
four-component vectors with lower indices, $v \equiv v_a$, as $v^{\
\prime} = U \ v$, or $v^{\ \prime}_a = U_a^{\quad b} \ v_b$. The
second equation in (\ref{usp4g}) expresses the invariance of the
symplectic product $u^T \ \Omega \ v \equiv u_a \ \Omega^{ab} \ v_b$:
\be
(u^T \ \Omega \ v)^{\ \prime} = u^T \ ( U^T \ \Omega \ U ) \ v =
u^T \ \Omega \ v \, .
\ee
We can then define symplectic vectors with upper indices by:
\be
u^a \equiv \Omega^{ab} \ u_b \, ,
\ee
and write the symplectic product as:
\be
u^T \ \Omega \ v = u_a \ \Omega^{ab} \ v_b 
= - u^b \ v_b  \equiv - u\, v \, .
\ee
As a short-hand notation for the symplectic product, we adopt the
NW-SE convention on the contraction of symplectic indices:
\be
u \, v \equiv u^b \ v_b \, .
\ee
Our convention on how to lower the indices is set by defining the
inverse symplectic metric with lower indices:
\be
u_a \equiv \Omega_{ab} \ u^b \qquad \Leftrightarrow \qquad
\Omega_{ab} \ \Omega^{bc} = \Omega^{cb} \ \Omega_{ba} 
= \delta_a^c \, .
\ee

The generators $a \equiv a_a^{\quad b}$ of the $USp(4)$ Lie algebra,
defined by $U = \exp (a)$, satisfy:
\beq 
a = - a^\dagger \, ,
\qquad 
a^T \ \Omega + \Omega \ a = 0 \, .
\label{eq:usp4algebra} 
\eeq 
Exploiting the isomorphism $Spin(5) \sim USp(4)$, the ten $USp(4)$
generators $a$ can be identified with those of $Spin(5)$, whose
elements $(\Gamma_i)_a^{\quad b}$ satisfy the Euclidean Clifford
algebra
\beq 
\{ {\Gamma_i} , {\Gamma_j} \}_a^{\quad b} = 2 \ \delta_{i j} 
\ \delta_a^b \, ,
\label{euclif}
\eeq 
where $i, j = 1, \dots, 5$ are the indices~\footnote{We are entitled
to use indifferently upper or lower indices of type $i, j = 1, \dots,
5$; we thus move them around freely for notational convenience.} of
$Spin(5)$ and $a, b = 1, \dots, 4$ are those of $USp(4)$. Notice that
we can define antisymmetric $\Gamma$ matrices with upper $USp(4)$
indices by:
\be
\Gamma_i^{ac} \equiv \Omega^{ab} \ (\Gamma_i)_b^{\quad c} \, ,
\qquad
\Gamma_i^{ac}= - \Gamma_i^{ca} \, .
\ee
Then, the antisymmetric two-index representation of $Spin(5)$ of
dimension 10, with generators
\be
\Gamma_{i j} \equiv {\Gamma_i \Gamma_j - \Gamma_j
\Gamma_i \over 2} \, ,
\ee
satisfies the algebra of eq.~(\ref{eq:usp4algebra}) automatically.
This means that the generic tensor $S_{a_1 \dots a_n}$ transforms
under $USp(4)$ as
\beq
S^\prime_{a_1\dots a_n} = 
U_{a_1}^{\quad b_1} \dots U_{a_n}^{\quad b_n} 
\ S_{b_1 \dots b_n} \, , 
\qquad 
U_{a}^{\quad b} = \exp \left( \frac12 \ \alpha^{ij} \ \Gamma_{i j}
\right)_{a}^{\quad b} \, ,
\eeq
where the $\alpha^{ij}$ are real coefficients. 

The generic $USp(4)$ algebra element with two indices can now be
decomposed as
\beq
S_a^{\quad b} = S \ \delta_a^b + S^i \ (\Gamma_i)_a^{\quad b} +
S^{ij} \ (\Gamma_{ij})_a^{\quad b} \, , \label{eq:decomp}
\eeq
where
\beq
S = {1 \over 4} \, S_a^{\quad a} \, ,
\qquad 
S^i = {1 \over 4} \, S_a^{\quad b} \, (\Gamma_i)_b^{\quad a} \, ,
\qquad 
S^{ij} = - {1 \over 2} \, S_a^{\quad b} \, (\Gamma_{ij})_b^{\quad a} \, .
\eeq
The last two relations link the {\bf 5} and the {\bf 10} of $USp(4)$ to
those of $SO(5)$.

An explicit representation of the $(\Gamma_i)_a^{\quad b}$ appearing
in (\ref{euclif}), useful for discussing the gaugings and the
Scherk--Schwarz twists in ${\cal N}=4$, $\De5$ supergravity is:
\be
\Gamma_{i=1\dots 3}=-\sigma_2\otimes\sigma_{i=1\dots 3} \, , 
\qquad
\Gamma_{i=4}=\sigma_1\otimes\id \, ,
\qquad
\Gamma_{i=5}=\sigma_3\otimes\id \, .
\ee
A convenient embedding of $U(1) \times SU(2) \subset USp(4)$ is:
\be
T_1 \equiv  - i \ \Gamma_{45} = - \sigma_2 \otimes \id \, ,
\quad
T_2 \equiv  - i \ \Gamma_{53} = \sigma_1 \otimes \sigma_3 \, ,
\quad
T_3 \equiv  - i \ \Gamma_{34} = \sigma_3 \otimes \sigma_3 \, ,
\label{eq:su2u1}
\ee
\be
Y \equiv  -i \ \Gamma_{12} = \id \otimes \sigma_3 \, ,
\quad 
[T_i , T_j ] = 2 \ i \ \epsilon_{i j k} \ T_k \, , 
\quad 
[ Y , T_i ] = 0 \, .  \label{eq:su2u1bis}
\ee
The fields of $\Ne4$, $\De5$ supergravity fall only in the $\bf 1$,
$\bf 4$ and $\bf 5$ irreducible representations of $USp(4)$: we
denote them here by the generic symbols $\phi$, $\chi_a$ and $A_{a
b}$, with $A_{a b}$ antisymmetric and $\Omega$-traceless. If we
parametrize the generic transformation by
\be 
U = U_a^{\ b} = \exp \Bigl( {i \over 2} \ \alpha^{ij} \ T_{ij} 
\Bigr)_a^{\ \ b} 
= 1 + {i \over 2} \ \alpha^{ij} \ {T_{ij}}_a^{\ \ b} + \ldots \, ,
\ee
\be
{T_{ij}}_a^{\ b}=-i {\Gamma_{ij}}_a^{\ b} \, ,
\ee
the non-trivial field transformations read
\bea 
& \chi_a \rightarrow \chi_a^\prime
=\exp \left( {i \over 2} \ \alpha^{ij} \ T_{ij} 
\right)_a^{\ b} \chi_b \, , &
\\ 
& A_{a b} \rightarrow A_{a b}^\prime = 
\exp \left( {i \over 2} \ \alpha^{ij} \ T_{ij} 
\right)_a^{\ c} \exp \left( {i \over 2} 
\ \alpha^{hk} \ T_{hk} \right)_b^{\ d} A_{c d} \, . &
\label{eq:5trans1}
\eea
Using the decomposition in \eq{eq:decomp} and the Clifford algebra in
\eq{euclif}, the infinitesimal transformation for $A_{a b}$ in the
$\mathbf{5}$ of $USp(4)$ reads
\beq 
\delta A^i = \alpha^{hk} \l
(\delta_{ih}\delta_{jk}-\delta_{ik}\delta_{jh} \r ) A^j 
=i \ \alpha^{hk} \ (\op_{hk})_{ij} A^j   \, ,
\eeq
and is just the transformation law for the fundamental
representation of $SO(5)$. We can thus exploit this further
isomorphism to rewrite the finite transformation of
eq.~(\ref{eq:5trans1}):
\beq
A_i\rightarrow {A_i}^\prime=\exp( i \ \alpha^{hk} \ \op_{hk} 
)_i^{\ j} A_{j} \, ,
\eeq
where $\op_{hk}$ are the 10 generators of $SO(5)$. Hence the $U(1)
\times SU(2)$ generators in eqs.~(\ref{eq:su2u1}) and
(\ref{eq:su2u1bis}) correspond to the $SO(2) \times SO(3)$ subgroup of
this $SO(5)$. A generic field $A^i$ transforming in the $\bf 5$ of
$USp(4)$ can then be decomposed as
\be
A^{\pm} \equiv {A^1 \mp i A^2 \over \sqrt2} \equiv 
A^{\frac{1\mp i\,2}{\sqrt2}} \sim {\bf 1}_\pm \, ,
\qquad A^{I=3,4,5} \sim {\bf 3}_0 \, ,
\ee
where ${\bf X}_{c}$ denotes a $\bf X$ of $SU(2)$ with $U(1)$ charge
equal to $c$, and we have introduced another short-hand notation to
identify the composition of a charged field.

Considering only two commuting generators of $U(1) \times SU(2)
\subset USp(4)$, $Y$ and $T_3$, we can write the explicit
transformation laws of the relevant fields as
\be
U = e^{{i \over 2} \l( \alpha_1 Y + \alpha_2 T_3 \r) } \, ,
\qquad
\chi_a \rightarrow (U_{4})_a^{\ b} \ \chi_b \, ,
\quad
\widetilde{A}_i \rightarrow (U_{5})_i^{\ j} \widetilde{A}_j \, ,
\label{eq:U}
\ee
where 
\be
U_4= {\rm diag} \left( 
e^{{i(\alpha_1+\alpha_2)\over 2}}, 
e^{ -{i(\alpha_1+\alpha_2)\over 2}},
e^{ {i(\alpha_1-\alpha_2)\over 2}}, 
e^{-{i(\alpha_1-\alpha_2)\over2}}
\right) \, ,
\ee
\be
U_5 = {\rm diag} \left( 
e^{i \alpha_1}, e^{-i \alpha_1}, e^{i \alpha_2}, 
e^{-i \alpha_2}, 1 \right) \, ,
\ee
\be
\widetilde{A}^T = \left(
A^{\frac{1-i \, 2}{\sqrt2}}, 
A^{\frac{1+i \, 2}{\sqrt2}},
A^{\frac{3-i \, 4}{\sqrt2}}, 
A^{\frac{3+i \, 4}{\sqrt2}}, 
A^{5} \right) = (A_1^+,
A_1^-, A_2^+, A_2^-, A_0) \, .
\label{eq:defvecss}
\ee
\subsection{Space-time and spinors}
\label{sec:spinors}
We choose a `mostly plus' metric:
\be
\eta_{AB} = {\rm diag}(-1,+1,+1,+1,+1) \, ,
\label{metric}
\ee
with Clifford algebra
\be
\left\{ \gamma^A \ , \ \gamma^B \right\} = - 2 \, \eta^{AB} \, .
\label{dirac}
\ee
Curved space-time indices are denoted with $M=(\mu,5)$, flat
tangent-space indices with $A=(\alpha,\widehat 5)$. The f\"unfbein is
$E_M^{\ A}$, its determinant $e_5 = \det \ E_M^{\ A}$. The total
antisymmetric tensor $\varepsilon^{MNPQR}$ is defined in such a way
that:
\beq 
\varepsilon^{MNPQR} = e_5 \ E_A^{\ M} E_B^{\ N} E_C^{\ P} 
E_D^{\ Q} E_E^{\ R} \
\varepsilon^{ABCDE} \, , \quad
\varepsilon^{\widehat0\widehat1\widehat2\widehat3\widehat5} = +1 \, ,
\eeq
\beq
\varepsilon^{\alpha\beta\gamma\delta \widehat 5} = 
\varepsilon^{\alpha\beta\gamma\delta} \, ,
\quad
\varepsilon^{\mu \nu \rho \sigma} = e_4 \ e_\alpha^\mu 
e_\beta^\nu e_\gamma^\rho e_\delta^\sigma \
\varepsilon^{\alpha \beta \gamma \delta} \, .
\eeq
Our explicit representation for the Dirac matrices is:
\bea 
\gamma^\alpha = \l ( \begin{array}{c c} 0&\sigma^\alpha \\ {\ov
\sigma}^\alpha&0 \end{array} \r )\,, \quad \gamma^{\widehat5} = 
\l ( \begin{array}{cc} -i&0 \\ 0&i \end{array} \r )\,, \quad 
\widehat \gamma = i \gamma^{\widehat5} =
\l ( \begin{array}{c c} 1&0 \\ 0&-1 \end{array} \r )\,,
\eea
\beq 
\sigma^{\alpha}=(-I,{\vec \sigma}) \, , \qquad {\ov \sigma}^{\alpha}=
(-I,-{\vec \sigma}) \, ,  
\eeq
with
\be
\gamma_M = E_M^{\ A} \gamma_A, \quad \gamma^{AB} =\frac12 
\left[\gamma^A, \; \gamma^B\right] = 2 \, \Sigma^{AB} \, ,
\ee
\be
\gamma^{ABCDE} = -\epsilon^{ABCDE} \, , \quad 
\gamma^{ABCD} = \epsilon^{ABCDE} \, \gamma_E \, , \quad 
\gamma^{ABC} = \epsilon^{ABCDE} \, \Sigma_{DE} \, .
\ee
Connections, covariant derivatives and curvature tensors are defined
as follows:
\begin{eqnarray}
&& \Gamma^M_{\ \ NR}=\frac12 G^{MS} \l ( \partial_N G_{RS}
+\partial_R G_{NS}-\partial_S G_{NR}\r ) \, , 
\nn \\ 
&& \omega^{\ \ AB}_M=2E^{N[A}\partial_{[N}E_{M]}^{\ B]}-E^{N[A}E^{R
B]}E_{MC}\partial_{R}E_N^{\ C} \, ,
\nn \\ 
&& D_M E^{\ A}_N=\partial_M E^{\ A}_N
-\Gamma^R_{\ \ MN} E^{\ A}_R-\omega_{M \ B}^{\
\ A} E^{\ B}_N=0 \, ,
\nn \\ 
&& R^{\quad \ AB}_{MN}=2\partial_{[M}\omega_{N]}^{\ \ AB}+2\omega^{\ \
AC}_{[N}\omega_{M]\ \ C}^{\ \ \ B} \, , 
\nn \\ 
&& R=R^{\quad \ AB}_{MN}E_A^{\ M} E^{\ N}_{B} \, ,
\nn \\ 
&& R^M_{\ \ NRS}=R_{RS}^{\quad BA}E^{\ M}_A E_{NB} = 2 \partial_{[R}
\Gamma^M_{\ \ S]N} + 2 \Gamma^M_{\ \ T[R} \Gamma^T_{\ \ S]N} 
\, , \nn
\end{eqnarray}
\begin{equation}
D_M\Psi = \partial_M\Psi +\frac12\omega_{MAB}\Sigma^{AB} \Psi \, , 
\qquad
[D_M,  D_N]\Psi = \frac12 R_{MNAB}\Sigma^{AB}\Psi \, ,
\end{equation}
where 
\be
A_{[M}B_{N]} \equiv {1 \over 2} (A_M B_N-A_N B_M) \, .
\ee

The $\De5$ charge conjugation matrix $C$ must obey, in any
conventions, the following two general properties:
\be
C \, \gamma^A \, C^{-1} = (\gamma^A)^T \, ,
\quad
C^T = - C \, .
\ee
We can then formulate the symplectic Majorana condition on spinors as
\be
\psi^a = C \, \ov{\psi_a}^T = C (\ov{\psi}^T)^a \, ,
\ee
where in writing the second equality we have exploited our previous
conventions for raising and lowering symplectic indices. From the
above conventions, the following hermiticity relations for symplectic
Majorana spinors follow:
\be
\ov{\psi}^a \, \gamma^{A_1} \ldots \gamma^{A_n} \, \chi_a =
- \ov{\chi}^a \, \gamma^{A_n} \ldots \gamma^{A_1} \, \psi_a =
 - (\ov{\psi}^a \, \gamma^{A_1} \ldots \gamma^{A_n} \, 
\chi_a)^\dagger \, .\label{eq:hermgamma}
\ee
Notice also that the symplectic metric acts like the charge 
conjugation matrix in the symplectic space of the $\Gamma_i$:
\beq
\Omega \, \Gamma_i \, \Omega^{-1}=\Gamma_i^T \, , 
\qquad \Omega=-\Omega^T \, ,
\eeq
so that:
\beq
\ov{\psi}^a ({\Gamma_{i_1}}\dots{\Gamma_{i_n}})_a^{\ b}  \chi_b=
-\ov{\chi}^a ({\Gamma_{i_n}}\dots{\Gamma_{i_1}})_a^{\ b}  \psi_b =
- [\ov{\psi}^a ({\Gamma_{i_1}}\dots{\Gamma_{i_n}})_a^{\ b}  
\chi_b]^\dagger \, .
\label{eq:hermGamma} 
\eeq
Then, all fermionic bilinears of the form $ \ov{\psi}^a
({\Gamma_{i_1}} \dots {\Gamma_{i_n}})_a^{\ b} \, \gamma^{A_1} \ldots
\gamma^{A_n} \, \chi_b$ are anti-hermitian.
%

%%%%%%%%%%%%%%%%%%%%%%%%%%%%%%%%%%%%%%%%%%%%%%%%%%%%%%%%%%%%%%%%%%%%
%%                          SECTION                               %%
%%%%%%%%%%%%%%%%%%%%%%%%%%%%%%%%%%%%%%%%%%%%%%%%%%%%%%%%%%%%%%%%%%%%

\section{Ungauged \boldmath{$\Ne4$}, \boldmath{$\De5$} supergravity} 
\label{app:5dlag}
In our conventions, and neglecting four-fermion terms that are not
relevant to the present work, the ungauged $\Ne4$, $\De5$
Lagrangian for pure supergravity \cite{cremmer, awatow} reads:
\bea 
e_5^{-1}{\cal L}&=& -R_5-\frac12\partial_M \phi \, \partial^M \phi
-\frac14 X^4 v_{MN}v^{MN}-\frac14 X^{-2} V^i_{MN} V^{MN}_i \nn \\ &&
-\frac18 e_5^{-1} \varepsilon^{MNRST} V_{MN}^i V_{RS}^i v_T - {i \over
  2} \ov \psi_M^a \gamma^{MNR} D_N \psi_{R \ a} + {i \over 2}
\ov\chi^a\gamma^M D_M \chi_a \nn \\ && -\frac{i}{8\sqrt2}\l ( X^{-1}
V_{MN}^i \Gamma_{i \, a}^{\quad \,b}+\frac{1}{\sqrt2} X^2 \delta_a^{\
  b} v_{MN} \r) \ov \psi^a_R
\gamma^{[R}\gamma^{MN}\gamma^{S]}\psi_{S\ b} \nn \\ &&
-\frac{i}{4\sqrt6}\l ( X^{-1} V_{MN}^i \Gamma_{i \, a}^{\quad \,
  b}-\sqrt2 X^2 \delta_a^{\ b} v_{MN} \r) \ov \psi^a_R
\gamma^{MN}\gamma^{R}\chi_{b} \nn \\ && +\frac{i}{24\sqrt2}\l ( X^{-1}
V_{MN}^i \Gamma_{i \, a}^{\quad \, b}-\frac{5}{\sqrt2} X^2 \delta_a^{\
  b} v_{MN} \r) \ov \chi^a \gamma^{MN} \chi_{b} \nn \\ &&
+\frac{i}{2\sqrt2}\partial_N \phi \, \ov \psi^a_M \gamma^N \gamma^M
\chi_a  \, ,
\label{eq:5dN4lag}
\eea
where:
\beq 
X = \exp\l(-\frac{\phi}{\sqrt6} \r) \, , \qquad V_{MN}^i = 2 \
\partial_{[M}V_{N]}^i \, , \qquad v_{MN} = 2 \ \partial_{[M}v_{N]} \, .
\eeq
Neglecting three-fermion terms, the supersymmetry transformation laws are:
\bea 
\delta E_M^{\ A} & = & {i \over 4} \ov \psi_M^a \gamma^A \epsilon_a \, ,
\nn \\ 
\delta V_M^i & = & \frac{i}{2\sqrt2}X \l (\ov \psi^a_M
+\frac{1}{\sqrt3}\ov \chi^a \gamma_M\r)\Gamma^{i \ \ b}_{\
a}\epsilon_b \, ,
\nn \\ 
\delta v_M & = & \frac{i}{4}X^{-2}\l ( \ov \psi_M^a
-\frac{2}{\sqrt3} \ov \chi^a \gamma_M\r)\epsilon_a \, , 
\nn \\ 
\delta \phi &=& \frac{i}{2\sqrt2}\ov \chi^a \epsilon_a \, ,
\nn
\\
\delta \psi_{M \, a} &=& D_M \epsilon_a -\frac{1}{12\sqrt2}\l (
\gamma_{M}^{\quad NR}+4\delta_{M}^{\ N}\gamma^R\r) \l (
X^{-1}V_{NR}^i\Gamma_{i \, a}^{\quad b}+
\frac{1}{\sqrt2}X^2\delta_a^{\ b} v_{NR}\r)\epsilon_b \, ,
\nn \\ 
\delta \chi_{a} &=&\frac{1}{2\sqrt2}\gamma^M\partial_M\phi
\,\epsilon_a +\frac{1}{4\sqrt6}\gamma^{MN}\l ( X^{-1}V_{MN}^i
\Gamma_{i \, a}^{\quad b}-\sqrt2 X^2\delta_a^{\ b} v_{MN}\r)
\epsilon_b \, .  \label{eq:5dsusy} 
\eea

%%%%%%%%%%%%%%%%%%%%%%%%%%%%%%%%%%%%%%%%%%%%%%%%%%%%%%%%%%%%%%%%%%%%
%%                          SECTION                               %%
%%%%%%%%%%%%%%%%%%%%%%%%%%%%%%%%%%%%%%%%%%%%%%%%%%%%%%%%%%%%%%%%%%%%

\section{The \boldmath{$\Ne4$}, \boldmath{$\De4$} reduced supergravity} 
\label{app:4dlagss}
We give here Lagrangian and transformation laws for the fully broken
$\De4$ effective theory, obtained via generalized reduction from the
$\Ne4$, $\De5$ ungauged theory. The unbroken theory corresponding to
the standard reduction can be easily extracted by setting the
Scherk--Schwarz twist $M=0$, while the partially broken theory can be
obtained by choosing $|m_1|\ne|m_2|=0$ in $M$. In the broken cases one
can move to the unitary gauge, removing the goldstinos according to
the standard procedure. Symplectic indices are suppressed according to
the NW--SE convention. Neglecting as before four-fermion terms, the
Lagrangian reads:
\beq
\cL^{SS}=\cL_{\rm bos}^{SS}+\cL_{\rm fer}^{SS} \, ,   
\label{eq:lag4dss}
\eeq
with
\bea
&& \hspace{-12pt} e_4^{-1} \cL_{\rm bos}^{SS} = 
\nn \\ &&
-R_4- {1 \over 2} { \partial_\mu t \partial^\mu t 
+ \partial_\mu \tau \partial^\mu \tau \over t^2}
- { \partial_\mu \varphi_0 \partial^\mu \varphi_0
+\widehat{D}_\mu \varphi_i \widehat{D}^\mu \varphi^i 
\over \varphi_0^2} 
\nn \\ &&
-\frac14  \, g_{IJ} \, \widehat{F}^I_{\mu\nu} 
\widehat{F}^{J \ \mu\nu} -\frac18 \, e_4^{-1} \theta_{IJ} \, 
\varepsilon^{\mu\nu\rho\sigma} \widehat{F}^I_{\mu\nu} 
\widehat{F}^{J}_{\rho\sigma} 
- {2 \over 3} i \ d_{\widehat{\imath}\, \widehat{\jmath} \,
\widehat{k}}\, M^{\widehat{k}}_{\ \ \widehat{l}} \,
\varepsilon^{\mu \nu \rho \sigma}  B_\mu^{\widehat{\imath}}
B_\nu^{\widehat{l}} B_{\rho \sigma}^{\widehat{\jmath}} \, , 
\\ \nn \\
&& \hspace{-12pt} e_4^{-1} \cL_{\rm fer}^{SS}= 
\nn \\ && 
-\frac12 e_4^{-1}
\eps^{\mu \nu \rho \sigma} \ov \eta_{\mu} {\widehat \gamma} \gamma_{\nu}
\l (\widehat{D}_{\sigma} + \frac12 \frac{\sqrt2}{t^{1/2} \varphi_0}
{\widehat\gamma} \gamma_{\sigma} M_4 \r )\eta_{\rho}
+\frac12 i \ov \chi \l ( \gamma^{\mu}
\widehat{D}_{\mu}+\frac{\sqrt2}{t^{1/2} \varphi_0} {\widehat\gamma} M_4
\r) \chi 
\nn \\ && 
+\frac34 i \ \ov \psi_y \l (\gamma^{\mu}
\widehat{D}_{\mu} + 2\frac{\sqrt2}{t^{1/2} \varphi_0} {\widehat \gamma}
M_4 \r) \psi_y
%\nn \\ &&
+ \frac32 \frac{\sqrt2}{t^{1/2} \varphi_0} \ov \psi_y \gamma^{\mu} M_4
\ \eta_{\mu} \nn \\ && - \frac18 \frac{t^{\frac12} \varphi_0}{\sqrt2}
\ov \eta_{\mu} \l [ \l ( \frac{\widehat{B}^i_{\rho \sigma}+A_{\rho
\sigma} \varphi^{i}}{\varphi_0}\Gamma_i+ \frac{b_{\rho \sigma}+A_{\rho
\sigma} \tau}{2t}\r){\widehat \gamma} -\frac{i}{2}A_{\rho \sigma} \r] \l (
e_4^{-1} \eps^{\mu \nu \rho \sigma} +2 i g^{\mu\rho}g^{\nu\sigma}
{\widehat \gamma}\r)\eta_{\nu} \nn \\ && -\frac14 \frac{t^{\frac12}
\varphi_0}{\sqrt2} \ov \psi_y \l [ \l ( \frac{\widehat{B}_i^{\mu
\nu}+A^{\mu \nu} \varphi_{i}}{\varphi_0}\Gamma_i+ \frac{b^{\mu
\nu}+A^{\mu \nu} \tau}{2t}\r) -\frac{3i}{2}A^{\mu \nu}{\widehat \gamma}
\r] \Sigma_{\mu \nu} \psi_y \nn \\ && + \frac{i}{12} \frac{t^{\frac12}
\varphi_0}{\sqrt2} \ov \chi \l [ \l ( \frac{\widehat{B}_i^{\mu
\nu}+A^{\mu \nu} \varphi_{i}}{\varphi_0}\Gamma_i- \frac52\frac{b^{\mu
\nu}+A^{\mu \nu} \tau}{t}\r) +\frac{3i}{2}A^{\mu \nu}{\widehat \gamma}
\r]\Sigma_{\mu \nu} \chi \nn \\ && -\frac{1}{8}\frac{t^{\frac12}
\varphi_0}{\sqrt2} \ov \psi_y \l [ \l ( \frac{\widehat{B}^i_{\rho
\sigma}+A_{\rho \sigma} \varphi^{i}}{\varphi_0}\Gamma_i+ \frac{b_{\rho
\sigma}+A_{\rho \sigma} \tau}{2t}\r){\widehat \gamma} +\frac{3i}{2}A_{\rho
\sigma} \r] \gamma^{\mu} \gamma^{\rho\sigma}\eta_{\mu} \nn \\ &&
-\frac{i}{2\sqrt3} \frac{t^{\frac12} \varphi_0}{\sqrt2}\ov \eta_{\mu}
\l ( \frac{\widehat{B}^i_{\rho \sigma}+A_{\rho \sigma}
\varphi^{i}}{\varphi_0}\Gamma_i- \frac52\frac{b_{\rho \sigma}+A_{\rho
\sigma} \tau}{t}\r) \Sigma^{\rho \sigma}\gamma^{\mu} \chi \nn \\ &&
-\frac{1}{2\sqrt3} \frac{t^{\frac12} \varphi_0}{\sqrt2}\ov \psi_y \l (
\frac{\widehat{B}_i^{\mu \nu}+A^{\mu \nu}
\varphi_{i}}{\varphi_0}\Gamma_i- \frac52\frac{b^{\mu \nu}+A^{\mu \nu}
\tau}{t}\r) {\widehat \gamma} \Sigma_{\mu \nu} \chi \nn \\ & &
+\frac{i}{4}\ov \eta_{\mu} \l (\frac{\widehat{D}_{\sigma}
\varphi^{i}}{\varphi_0}\Gamma_{i}+\frac{\partial_\sigma \tau}{2t} \r)
e_4^{-1}\eps^{\mu \nu \rho \sigma} \gamma_{\nu} \eta_{\rho}
+\frac{1}{8} \ov \psi_y \l (\frac{\widehat{D}_{\mu}
\varphi^{i}}{\varphi_0}\Gamma_{i}+\frac{\partial_{\mu} \tau}{2t} \r)
{\widehat \gamma} \gamma^{\mu} \psi_y \nn \\ & & -\frac{1}{12} \ov \chi \l
(\frac{\widehat{D}_{\mu}
\varphi^{i}}{\varphi_0}\Gamma_{i}-\frac52\frac{\partial_{\mu} \tau}{t}
\r) {\widehat \gamma} \gamma^{\mu} \chi
%\nn \\  & & 
-\frac{i}{2} \ov \psi_y \l (\frac{\widehat{D}_{\nu} (\varphi^{i}
  \Gamma_{i}-i\varphi_0 {\widehat \gamma})}{\varphi_0}
+\frac{\partial_{\nu} (\tau -i t {\widehat \gamma})}{2t} \r)
\gamma^{\mu}\gamma^{\nu} \eta_{\mu} \nn \\ && +\frac{i}{2\sqrt3}\ov
\eta_{\mu} \l (\frac{\partial_{\nu} (t +i \tau {\widehat \gamma})}{t}-
\frac{\widehat{D}_{\nu} (\varphi_0+i\varphi^{i} \Gamma_{i} {\widehat
    \gamma})}{\varphi_0} \r) \gamma^{\nu}\gamma^{\mu}\chi
%\nn \\ && 
+\frac{i}{\sqrt3} \ov \psi_y \l (\frac{\widehat{D}_{\mu}
\varphi^{i}}{\varphi_0}\Gamma_{i}-\frac{\partial_{\mu} \tau}{t} \r)
\gamma^{\mu} \chi \, . \nn \\
& & 
\eea
The explicit expressions for the covariant derivatives and the
four-dimensional fields are given in the text. The corresponding
supersymmetry transformations are, up to three-fermion terms:
\bea 
\delta e_\mu^\alpha &=& \frac{i}{4} \ov \eta_\mu \gamma^\alpha 
\epsilon 
%-\frac{1}{8}\ov \psi_y {\widehat \gamma} \gamma^{\alpha}_{\ \beta} 
%\epsilon \ e_{\mu}^{\beta} 
\, , \nn \\
\frac{\delta t}{t} &=& \frac{1}{4} \l ( \ov \psi_y {\widehat \gamma}
+{2i\over \sqrt3}\ov \chi \r ) \epsilon \, , 
\qquad 
\frac{\delta \tau}{t} = \frac{i}{4} \l ( \ov \psi_y
+{2i\over \sqrt3}\ov \chi {\widehat \gamma} \r ) \epsilon \, , 
\nn \\
\frac{\delta \varphi_0}{\varphi_0} &=& \frac{1}{4} \l ( 
\ov \psi_y {\widehat \gamma}-{i\over \sqrt3}\ov \chi \r ) 
\epsilon \, , 
\qquad 
\frac{\delta \varphi^i}{\varphi_0} = \frac{i}{4} \l ( \ov 
\psi_y-{i\over \sqrt3}\ov \chi {\widehat \gamma} \r ) 
{\Gamma^{\,i}}\,\epsilon \, , 
\nn \\
\frac{\varphi_0 t^{1/2}}{\sqrt2} \delta A_\mu &=& \frac{1}{4} 
\ov \eta_\mu {\widehat \gamma} \epsilon \, , \nn \\
\frac{\varphi_0 t^{1/2}}{\sqrt2} \delta b_\mu &=& 
\frac{i}{4} \ov \eta_\mu \l ( t+i\tau {\widehat \gamma} \r )
\epsilon + \frac{t}{8} \l ( \ov \psi_y {\widehat \gamma} -
\frac{4i}{\sqrt3} \ov \chi \r) \gamma_\mu \epsilon \, , 
\nn \\
\frac{\varphi_0 t^{1/2}}{\sqrt2} \delta B^i_\mu &=& 
\frac{i}{4} \ov \eta_\mu \l ( \varphi_0 {\Gamma^{\,i}}
+i\varphi^i {\widehat \gamma} \r )\epsilon +\frac{\varphi_0}{8}  
\l ( \ov \psi_y {\widehat \gamma} +\frac{2i}{\sqrt3} \ov \chi 
\r) \gamma_\mu {\Gamma^{\,i}}\,\epsilon \, , 
\nn \\
\delta \eta_{\mu} &=& \widehat{D}_\mu \epsilon  
+\frac{1}{2}\frac{\sqrt2}{t^{1/2}\varphi_0}{\widehat \gamma} \gamma_\mu
M_4 \ \epsilon
+ \frac{i}{2}{\widehat \gamma} 
\l ( {\widehat{D}_\mu \varphi^i \over \varphi_0}\Gamma_{i}
+\frac12 {\partial_\mu \tau \over t}\r)\epsilon  
\nn \\ &&
-\frac18 \frac{\varphi_0 t^{1/2}}{\sqrt2} \l ( 
\gamma_\mu^{\ \nu \rho}+2\delta_\mu^\nu \gamma^\rho\r)
\l (\frac{\widehat{B}_{\nu\rho}^i+\varphi^i A_{\nu\rho}}{\varphi_0} 
\Gamma^i+\frac{b_{\nu\rho}+\tau A_{\nu\rho}}{2t}-\frac{i}{2}
{\widehat \gamma} A_{\nu\rho}\r)\epsilon 
\nn \\
\delta \psi_{y} &=& i \frac{\sqrt2}{t^{1/2}\varphi_0} M_4 \ \epsilon
+\frac{i}{3} \l ( \frac{\partial_\mu(t
-i{\widehat \gamma} \tau)}{2t}+\frac{\widehat{D}_\mu (\varphi_0
-i{\widehat \gamma} \varphi^i \Gamma_i)}{\varphi_0}\r) 
{\widehat \gamma} \gamma^\mu \epsilon 
\nn \\
&&+\frac{i}{12}\frac{\varphi_0 t^{1/2}}{\sqrt2}{\widehat \gamma}
\gamma^{\mu\nu}\l (\frac{b_{\mu\nu}+\tau A_{\mu\nu}}{2t}+
\frac{\widehat{B}_{\mu\nu}^i+\varphi^i A_{\mu\nu}}{\varphi_0}\Gamma^i
-\frac32 i {\widehat \gamma} A_{\mu\nu}\r) \epsilon \, , 
\nn \\
\delta \chi &=& \frac{1}{2\sqrt3} \l ( \frac{\partial_\mu( t
-i{\widehat \gamma} \tau)}{t}-\frac{\widehat{D}_\mu ( \varphi_0
-i{\widehat \gamma} \varphi^i \Gamma_i)}{\varphi_0}\r) \gamma^\mu 
\epsilon \nn \\
&&+\frac{1}{4\sqrt3}\frac{\varphi_0 t^{1/2}}{\sqrt2}\l (
\frac{\widehat{B}_{\mu\nu}^i+\varphi^i A_{\mu\nu}}{\varphi_0}\Gamma^i 
- \frac{b_{\mu\nu}+\tau A_{\mu\nu}}{t}\r)\gamma^{\mu\nu} \epsilon \, , 
\label{eq:susytransf4d} 
\eea
where the infinitesimal parameter $\epsilon$ has been rescaled as
\be
\epsilon_{\De5}=\rho^{-1/4} \ \epsilon_{\De4} \, ,
\ee
and its covariant derivative is defined as
\beq
\widehat{D}_\mu \epsilon_a = \partial_\mu \epsilon_a - i A_\mu \
(M_4)_a^{\ b} \ \epsilon_b \, . \nn \\ 
\eeq
%

%%%%%%%%%%%%%%%%%%%%%%%%%%%%%%%%%%%%%%%%%%%%%%%%%%%%%%%%%%%%%%%%%%%%
%%                       BIBLIOGRAPHY                             %%
%%%%%%%%%%%%%%%%%%%%%%%%%%%%%%%%%%%%%%%%%%%%%%%%%%%%%%%%%%%%%%%%%%%%

%
\end{document}